\documentclass[12pt,preprint]{aastex}

\slugcomment{Submitted to Astrophys. J.}

\shorttitle{Cepheid Distance Scale}
\shortauthors{Barnes et al.}

\begin{document}

\title{A Bayesian Analysis of the Cepheid Distance Scale}

\author{Thomas G. Barnes III}
\affil{McDonald Observatory, University of Texas at Austin, Austin, TX
78712}
\email{tgb@astro.as.utexas.edu}

\author{W. H. Jefferys}
\affil{Department of Astronomy, University of Texas at Austin, Austin, TX
78712}
\email{bill@astro.as.utexas.edu}

\author{J. O. Berger and Peter J. Mueller\altaffilmark{1}}
\affil{Institute for Statistics and Decision Sciences, Duke University,
Durham, NC 27708}
\altaffiltext{1}{Now at M. D. Anderson Cancer Center, Houston,
TX.}

\author{K. Orr\altaffilmark{2}}
\affil{Department of Astronomy, University of Texas at Austin, Austin, TX
78712}
\altaffiltext{2}{Now at Baylor University, Waco, TX.}

\and

\author{R. Rodriguez\altaffilmark{3}}
\affil{Department of Astronomy, University of Texas at Austin, Austin, TX
78712}
\altaffiltext{3}{Now at Northwestern University, Evanston, IL}

\begin{abstract}

We develop and describe a Bayesian statistical analysis to solve the surface brightness equations for
Cepheid distances and stellar properties. Our analysis provides a mathematically rigorous and objective
solution to the problem, including immunity from Lutz-Kelker bias.  We discuss the choice of priors, show
the construction of the likelihood distribution, and give sampling algorithms in a Markov Chain Monte Carlo
approach for efficiently and completely sampling the posterior probability distribution.  Our analysis
averages over the probabilities associated with several models rather than attempting to pick the `best
model' from several possible models. Using a sample of thirteen Cepheids we demonstrate the method. We
discuss diagnostics of the analysis and the effects of the astrophysical choices going into the model. We
show that we can objectively model the order of Fourier polynomial fits to the light and velocity data. 
By comparison with theoretical models of Bono {\em et al.} (2001) we find that EU Tau and SZ Tau are
overtone pulsators, most likely without convective overshoot. The period-radius and
period-luminosity relations we obtain are shown to be compatible with those in the recent
literature. Specifically, we find
$log(<R>) = 0.693{(\pm 0.037)}(log(P)-1.2)+2.042{(\pm 0.047)}$ and $<M_v> = -2.690{(\pm
0.169)}(log(P)-1.2)-4.699{(\pm 0.216)}$. 

\end{abstract}


\section{Introduction}

Astrophysics is replete with statistical analyses of research problems.  Overwhelmingly the 
approach applied is frequentist. However, in many problems there are advantages to a Bayesian 
approach. It is coherent and systematic, and in principle it is straightforward to write down all of the
elements of even a complex mathematical and statistical model that fairly captures our understanding of
the problem. The Bayesian approach is also well-suited to objective selection of an empirical model, such
as a Fourier polynomial of unknown order to represent the pulsations of a variable star, from many possible
polynomials competing for our attention. It addresses the errors-in-variables problem and the
proper handling of errors in the data in a natural way.  In this paper we apply Bayesian statistics to a
problem of considerable astrophysical interest: the distance scale of Cepheid variables as determined 
through the surface brightness technique. 

The surface brightness technique is widely applied to determination of the Cepheid period-luminosity and 
period-radius relations.  Applications of the method have improved steadily since it was first discussed
by  Barnes {\em et al.} (1977) \nocite{bde1977} but are still not mathematically rigorous and objective.
The essence of the  technique is to infer the stellar angular diameter from an appropriate magnitude and
color index pair, compare this angular diameter throughout the pulsation cycle with linear displacements
of the stellar atmosphere obtained by integrating the radial velocity curve, and extract the mean stellar
radius and the stellar distance. To our knowledge none of the published applications objectively selects,
as part of the problem, a model for the radial velocity curve so that it may be integrated.  This is
straightforward in our Bayesian approach.  Secondly, none of the published applications handle
propagation of the radial velocity uncertainties through the integration in a mathematically rigorous
way.  The analysis in our work does so.  Finally, both the angular diameters and the linear displacements
have error; not all published applications treat this {\em errors-in-variables} problem correctly. Again,
our approach does this naturally.

A second reason to apply rigorous analysis to the surface brightness determination of Cepheid 
distances is that the Cepheid period-luminosity (PL) relation provides the foundation for the extragalactic
distance scale and all conclusions that derive from it. In their final report on the Hubble Space
Telescope  Key Project on $H_0$, Freedman {\em et al.} (2001) \nocite{freedman} list several difficulties 
with the Cepheid PL relation, among which is the following ``... an accurate geometric calibration 
of the PL relation, at any given metallicity, has not yet been established.''  The surface brightness
method has recently been calibrated geometrically through interferometric angular diameters of Cepheids
(Nordgren {\em et al.} 2002). Strong confidence in that calibration arises from its prediction of the
same parallax for $\delta$ Cep, within $1\%\pm4\%$, as obtained using the Hubble Space Telescope (Benedict
{\em et al.} 2002). The Nordgren {\em et al.} (2002) surface brightness calibration permits Cepheid
distances to be determined that are very close to, if not fully, geometric and that are fully independent
of all other astronomical distance scales.  In this work we apply that calibration to a modest selection
of Cepheids to gauge its effect upon the Cepheid distance scale.

\section{The Surface Brightness Relationship}

The visual surface brightness parameter $F_v$ was introduced by Barnes \& Evans
(1976) \nocite{be76} as
\begin{equation}
F_v=4.2207-0.1V{\rm _0}-0.5\log\phi \label{eq:fparameter}
\end{equation}
\noindent
and also,
\begin{equation}
F_v=\log{T_e}+0.1BC, \label{eq:eq.2}
\end{equation}
\noindent
where $V{\rm _0}$ is the stellar visual magnitude corrected for interstellar
extinction, $\phi$ is the stellar angular diameter expressed in milliarcseconds, $T_e$
is the effective temperature and $BC$ is the bolometric correction. In that paper and
in Barnes, Evans, \& Parsons (1976) \nocite{bep76} they showed that $F_v$ is tightly
correlated with Johnson color index
$(V-R){\rm _0}$ for a very wide range of stellar types.  Specifically for Cepheids,
Barnes {\em et al.} (1977) \nocite{bde1977} demonstrated a linear relationship 
\begin{equation}
F_v=A+B\left(V-R\right){\rm _0} \label{eq:basiceq}
\end{equation}
\noindent 
and used this relation to infer a distance scale for Cepheids. The parameters in Eq.(\ref{eq:basiceq}),
called the visual surface brightness relation, and its extension to other photometric bands have been
addressed over time by many authors. Discussions of previous work have been given by Fouqu\'{e} \& Gieren
(1997)\nocite{fg97} and by Nordgren {\em et al.} (2002)\nocite{nordgren2002}. 

Because there had been no high-quality angular diameters of Cepheids until very
recently, all attempts to determine the parameters of Eq.(\ref{eq:basiceq}) have been
indirect.  One approach has been to assume that the Cepheid surface brightness relation
is the same as that for nonvariable giants and supergiants with measured angular
diameters, for which Eq.(\ref{eq:basiceq}) can be calibrated through
Eq.(\ref{eq:fparameter}) (Welch 1994, \nocite{w94} Fouqu\'{e} \& Gieren
1997\nocite{fg97}). Another method has used the Cepheid temperature and bolometric correction
scales to calibrate Eq.(\ref{eq:basiceq}) through Eq.(\ref{eq:eq.2}), {\em e.g.} Gieren,
Barnes, \& Moffett (1993)\nocite{gbm93}. Still another approach used model atmosphere
calculations for Cepheids to establish the dependence, {\em e.g.}, Hindsley \& Bell (1986)
\nocite{hb86} and Gieren, Barnes, \& Moffett (1993)\nocite{gbm93}. Because these approaches established
the surface brightness relation indirectly, they are open to criticism that the resultant distance scale
may be adversely affected by the assumptions made.

Recently Nordgren {\em et al.} (2002)\nocite{nordgren2002}, used interferometrically determined angular
diameters of Cepheid variables to establish the parameters of Eq.(\ref{eq:basiceq}). 
Fifty-nine interferometrically measured angular diameters of the Cepheids $\eta$ Aql, $\delta$
Cep and
$\zeta$ Gem yielded
\begin{equation}
F_v=3.941({\pm 0.004})-0.368({\pm 0.007})\left(V-R\right){\rm _0} \label{eq:nordgren}
\end{equation}
\noindent
Because Eqs.(\ref{eq:nordgren}) is based upon measured angular diameters, the surface
brightness technique may finally fulfill its original promise: determination of essentially
geometric distances to Cepheids.  We have adopted Eq.(\ref{eq:nordgren}) for
the analyses in this paper.

\section{Cepheid distance determination}

As developed by Barnes {\em et al.} (1977) \nocite{bde1977} and extended by Gieren,
Barnes, \& Moffett (1993)\nocite{gbm93}, the surface brightness relation permits an independent distance
scale to be determined for Cepheids.  Those works may be consulted for full discussion; here we give a
synopsis of the method to lay the foundation for our Bayesian analysis.

At each time $t$ in the pulsation of the Cepheid, Eqs.(\ref{eq:fparameter})
and (\ref{eq:basiceq}) may be combined to obtain the angular diameter variation of
the star, $\phi(t)$, 
\begin{equation}
4.2207-0.1V{\rm _0}(t)-0.5\log\phi(t)=A+B\left(V-R\right){\rm _0}(t)
\label{eq:combined}
\end{equation}
\noindent
In addition we infer the Cepheid's linear radius variation ${\Delta}R(t)$ about the mean
radius from an integration of the radial velocity curve, $V_r(t)$, 
\begin{equation}
{\Delta}R(t)=-\int{p\left(V_r(t)-{V_\gamma}\right)dt}
\end{equation}
\noindent
where the factor $-p$ converts observed radial velocity to the star's pulsational
velocity and $V_\gamma$ is the center-of-mass radial velocity of the star.  Integration of
the discrete radial velocity data requires that the velocity variation be appropriately modeled. We 
follow convention and model the velocity variation as a Fourier series of order $M$.  However, unlike the
conventional approach, our Bayesian analysis permits $M$ to be determined as part of the problem, rather
than being selected by {\em ad hoc} means. Indeed, we will go further and average over the best models 
in an optimal way, thus improving  the results over simple selection of the ``best model,'' when more 
than one model is in the running.  We will similarly model the magnitude variation as a Fourier series of
order $N$ in order to obtain mean absolute magnitudes.

The distance $s$ and the mean angular diameter $\phi_0$ may be inferred from the solution to
\begin{equation}
4.2207-0.1V{\rm _0}(t)-0.5\log(\phi_0+2000\Delta R(t)/s)=A+B\left(V-R\right){\rm _0}(t).
\label{eq:combined2}
\end{equation}
 \noindent
where $\phi_0$ is in milliarcseconds, $s$ is in parsecs, and ${\Delta}R(t)$ is in AU.  (The factor $2000$
converts radius to diameter and arcseconds to milliarcseconds.) 

\section{Bayesian analysis}

We have adopted a Bayesian approach to solve for stellar distance and radius through the
surface brightness technique. This has a number of advantages. It is coherent and systematic,
and in principle it is straightforward to write down all of the elements of even a complex
mathematical and statistical model that fairly captures our understanding of the problem. The
Bayesian approach we will use is especially well-suited to the situation at hand, where
several empirical models for the pulsations of the star (represented as Fourier polynomials
of indeterminate order) are competing for our attention, and we need to choose the best
representation in a reasonably objective way. Thus, we are faced with a {\em
model selection} or {\em model averaging} situation; Bayesian analysis can be used
to approach such problems in a systematic way (Hoeting {\em et al.} 1999). \nocite{hoe99}
A good introduction to the basic ideas can be found in Loredo (1990).  \nocite{lor90}

As usual in Bayesian analysis, we consider the possible {\em states of
nature} $x$, and encode our uncertainty about the states of nature, before
looking at the data, in the form of a prior distribution $p(x)$ on the
states of nature. The states of nature include: the different possible
models we are considering, the parameters of each model ({\em e.g.}, Fourier
coefficients and variances peculiar to each model), parameters that are
common across models ({\em e.g.}, the parallax of the star, its linear
diameter, and an unknown phase factor representing any systematic
difference in phase between the photometry and the velocity data), and perhaps
other variables and hyperparameters introduced to complete the model.

This {\em prior} distribution is multiplied by the {\em likelihood}, which is
the probability $p(d|x)$ of obtaining the particular data $d$ we have
observed, conditioning on the data, as a function of the {\em states of nature} $x$.
This leads to the {\em Bayesian paradigm}, the posterior probability is proportional to
prior times likelihood:
\begin{equation}
p(x|d)\propto p(x) \times p(d|x) \label{eq:unnorm}
\end{equation}
As always in Bayesian analysis, the data, once observed, are conditioned upon and {\em
fixed}, and the resulting unnormalized {\em posterior probability distribution} is a
function only of the parameters that represent the states of nature. 

It can be argued that (almost) all Bayesian inference amounts to evaluating integrals with
respect to the posterior distribution $p(x|d)$. This includes posterior means for point
estimates, posterior predictive integrals for prediction, and more. However, in practice it
is usually impossible to do the integrals analytically. During the past decade and a half, a
different approach has been developed in which the {\em unnormalized} posterior distribution
Eq.(\ref{eq:unnorm}) can be used directly to generate a large Monte Carlo sample from the
posterior distribution. This is  implemented by computer simulation. Once we have this
sample, we can then derive all our  inferences from the sample. Thus, we can derive means,
medians and variances of  a parameter  of interest as sample means, medians, and variances of
the parameter in the sample, and can  also plot approximations to the posterior distributions
of particular parameters as histograms of the sample restricted to the particular parameters
in question. In principle, the sample encodes any posterior inference that
we might be interested in making.

A number of techniques can be used to generate this sample. The most common are Gibbs sampling
and Metropolis-Hastings sampling. These go more generally under the name of Markov Chain
Monte Carlo (MCMC), and can be used either separately or in combination to generate the
sample as a realization of a Markov chain whose stationary distribution is the
target posterior distribution (Smith \& Gelfand 1992)\nocite{smi92}. Using averages of the Monte Carlo samples we can approximate
any desired posterior integral by corresponding ergodic averages. For a practical implementation, an
initial burn-in is often discarded to achieve a better approximation. There are infinitely many possible
Markov chains that might be cooked up that will generate a sample from the posterior distribution. The
art is to devise a Markov chain that samples efficiently and completely, and this is more difficult.

We now describe details of the Gibbs sampling and Metropolis-Hastings algorithm for a
generic desired stationary distribution $p(x)$. In the application to posterior integration,
this is the posterior $p(x|d)$. To indicate a multivariate state vector we generally write
$(x,y)$, {\em etc.}

\subsection{Gibbs sampling}

The idea behind Gibbs sampling (Gelfand \& Smith 1990)\nocite{gf1990} is that, 
although we may not be able to draw a sample in all variables simultaneously from a high
dimensional distribution, we may nonetheless be able to draw samples from lower-dimensional 
conditional distributions on groups of variables, conditioned on the current values of all 
the other variables. (For a tutorial discussion  see Casella \& George 1992.) \nocite{cas92}. 
For example, we may have a distribution $p(x,y)$ from which we cannot directly draw a
random sample, but it may be possible to draw samples from the conditional
distributions $p(x|y)$ and $p(y|x)$. In this case, we can construct a Markov chain with the
desired joint distribution $p(x,y)$ as its stationary distribution as follows. Start from some
arbitrary point $(x_0,y_0)$ in the parameter space. Then draw $(x_1,y_1),
(x_2,y_2),\ldots$ successively by drawing $x_i$ from $p(x|y_{i-1})$ and
then drawing $y_i$ from $p(y|x_i)$. The result is a Markov chain whose
stationary distribution is $p(x,y)$. This generalizes to any number of
variables, and variables can be sampled in groups as well as singly,
{\em e.g.}, $x|y$ could be multivariate normal and the method would still work.

\subsection{Metropolis-Hastings sampling}

Sometimes it is not possible to sample from one or more of the
conditional distributions, as is done in Gibbs sampling. In such a
situation, we can resort to the somewhat more brute-force method of
Metropolis-Hastings sampling (Tierney 1994)\nocite{tierney}. (Again, for a tutorial,
see Chib \& Greenberg 1995.)\nocite{chi95} The idea is as follows. Suppose we wish to sample from the
probability  distribution $p(x)$, where again $x$ may be multivariate. We start at an 
arbitrary point $x_0$, and generate successive samples $x_1,x_2,\ldots$ by an 
appropriate rule. In particular, if the current position is $x$, we choose a 
{\em proposal value} $x^*$ by selecting a random variable from a more-or-less 
arbitrary {\em proposal distribution} $q(x^*|x)$. Compute the Metropolis-Hastings
ratio
\begin{equation}
\alpha=min \left[1,\frac{p(x^*)q(x|x^*)}{p(x)q(x^*|x)}\right]
\end{equation}
Then accept the move to $x^*$
with probability $\alpha$ and remain at $x$ with probability
$1-\alpha$. Thus, we either remain at the current point or move to a new
point, with a probability depending on the value of $\alpha$.

It is possible to mix Gibbs and Metropolis-Hastings steps in the same
calculation. Thus, if there is a subset of parameters for which we can
effectively sample from the conditional distributions, we may use Gibbs
sampling on that subset; the remaining variables may be sampled in
Metropolis-Hastings steps, in a so-called {\em Metropolis-within-Gibbs}
scheme.

A good deal of art and experience goes into constructing a good scheme. It is 
important that the Markov chain mix fast across all regions of the parameter space, 
that it move easily and freely from one region to another, that it not get ``stuck'' 
in one region of the parameter space for long, and so forth. Otherwise, it will take an
inordinately long time for the sampling scheme to draw a sample that is
useful for making inferences. Devising an effective sampling scheme
requires a good understanding of the problem. For example, if the
proposal distribution $q(x^*|x)$ is {\em close} to the actual
distribution $p(x^*)$ then the Metropolis-Hastings step will be almost a
Gibbs step, since the value of $\alpha$ will be close to 1 and the
proposed step will almost always be accepted. Also, it is often more
efficient to sample parameters in groups. If, for example, several
parameters are highly correlated, then approximate knowledge of the
correlation structure can be exploited to sample those parameters
efficiently. 

\subsection{Mathematical model and likelihood function}

We have unequally-spaced observations of velocity data
$U_i,i=1,\ldots,m$, and  photometric data consisting
of simultaneously-observed magnitude $V_j,j=1,\ldots,n$ and
color index
$C_j,j=1,\ldots,n$. We are given standard deviations
$\sigma_{U_i}$,
$\sigma_{V_j}$,
$\sigma_{C_j}$.  However, we are not very confident
of these numbers and take the variances of the data to be given by
$\sigma_{U_i}^2/\tau_U$, $\sigma_{V_j}^2/\tau_V$, $\sigma_{C_j}^2/\tau_C$,
where the {\em hyperparameters} $\tau_U$, $\tau_V$, $\tau_C$ are to be
estimated. Let $u_i$, $v_j$, and $c_j$
denote the unknown true velocity, magnitude, and color index,
respectively. Conditional on $u_i$, $v_j$ and $c_j$ we assume independent
normal distributions
\begin{eqnarray}
& U_i \sim N(u_i,\sigma^2_{U_i}/\tau_U)\nonumber\\  \label{eq:likelihood}
& V_j \sim N(v_j,\sigma^2_{V_j}/\tau_V)\\  
& C_j \sim N(c_j,\sigma^2_{C_j}/\tau_C)\nonumber  
\end{eqnarray}

The velocity $u$ and photometry $(v,c)$ are periodic functions of the time, and so are
functions of the pulsation phase $\theta$ where $0\le\theta<1$.  An obvious strategy is to
represent them as Fourier polynomials of some unknown or selectable order, resulting in a model
selection/averaging problem. We need to do this only for $u$ and $v$, since the colors $c$
are  mathematically related to $u$ and
$v$ through (Eq. \ref{eq:stefan}) below. $M$ and $N$ are the unknown order of the Fourier
polynomials for the $\mbox{\bf U}$ and
$\mbox{\bf V}$  data, respectively. The polynomials contain $2M+1$ and
$2N+1$ terms, respectively, including the leading constant terms. 
Thus we write
\begin{eqnarray*}
&\mbox{\bf u}=u_0+\mbox{\bf X}_u \mbox{\bf a}_u\\
&\mbox{\bf v}=v_0+\mbox{\bf X}_v \mbox{\bf a}_v
\end{eqnarray*}
where $u_0$ and $v_0$ are the mean velocity and apparent magnitude,
 $\mbox{\bf X}_u$ and
$\mbox{\bf X}_v$ are $(m \times 2M)$ and $(n \times 2N)$
design matrices consisting of sines and cosines of 
multiple angles, evaluated at the phases of the data, and $\mbox{\bf
a}_u$ and
$\mbox{\bf a}_v$ are vectors of Fourier coefficients.  Note that the
velocity data and photometric data are taken independently, so the phases
in general are different. Because the velocity and photometry data are
taken at different epochs, there may be an unknown phase error $\Delta\theta$ between
the two (due to imperfect knowledge of the period of the star and/or to change in the period between
the photometric and velocity epochs). 

Expressing Eq.(\ref{eq:combined2}) in our statistical nomenclature, we have the
nonlinear relationship
\begin{equation}
c_j=\frac{1}{B}\left(4.2207-0.1v_j-A-0.5\log\left(\phi_0+2000\Delta
R_i/s\right)\right) \label{eq:stefan}
\end{equation}
\noindent
where $(A,B)$ are known constants from Eq. (\ref{eq:nordgren}), $\phi_0$ and $s$ are the mean
angular diameter and the distance of the star (both to be estimated), and $\Delta R_i$, the linear radial
displacement, is calculated from the $\mbox{\bf a}_u$ by integrating the velocity term-by-term with respect
to the phase. This allows us to write down the likelihood function directly from Eqs.
(\ref{eq:likelihood}): 
\begin{eqnarray}
p(d|x)=\prod_{i=1}^m{\left[\sqrt{\tau_U}\exp\left(
-\frac{1}{2}\tau_U(U_i-u_i)^2\right)\right]}\nonumber \\
\times \prod_{j=1}^n{\left[\sqrt{\tau_V}\exp\left(
-\frac{1}{2}\tau_V(V_j-v_j)^2\right)\right]} \\
\times \prod_{j=1}^n{\left[\sqrt{\tau_C}\exp\left(
-\frac{1}{2}\tau_C(C_j-c_j)^2\right)\right]} \nonumber
\end{eqnarray}
Some of the parameters appear in the resulting likelihood function in awkward and
nonlinear ways through Eq. (\ref{eq:stefan}) that will make  straightforward Gibbs
sampling impossible. A suitably informed Metropolis-within-Gibbs scheme will be
needed.

\subsection{Priors} 

The posterior inference is summarized in the posterior distribution on
the following unknown parameters:
\begin{itemize}
\item[(1)] The orders of the Fourier models, $M$ (velocities) and $N$ ($V$ magnitudes).
\item[(2)] The precision hyperparameters $\tau_U$, $\tau_V$, $\tau_C$.
\item[(3)] The mean angular diameter $\phi_0$ and the unknown phase error
$\Delta\theta$. 
\item[(4)] The distance $s$.
\item[(5)] The intercepts $u_0$ and $v_0$ and the Fourier coefficients
$\mbox{\bf a}_u$,
$\mbox{\bf a}_v$.
\end{itemize}
The posterior distribution is also indirectly a function of additional 
hyperparameters introduced below, which define the prior probability model 
on the Fourier coefficients.

We expect the order of the models to be modest; we choose a uniform prior 
on the models $(M,N)$ up to some cut-off, and zero beyond.  The cut-off is
virtual in the sense that we ensure that $(M,N)$ are large enough to 
encompass all models with significant probability but not much larger to 
save computing time in the burn-in phase.

The precision hyperparameters $\tau_U$, $\tau_V$, $\tau_C$ are scale
variables. We use standard Jeffreys priors, proportional to the
reciprocal of the hyperparameter. Probably we could give them more
informative priors but it didn't seem necessary in this case. So
\begin{displaymath}
p(\tau_U)\propto 1/\tau_U,\; p(\tau_V)\propto 1/\tau_V,\;p(\tau_C)\propto
1/\tau_C
\end{displaymath}

We take the priors on $\Delta\theta$ and $\phi_0$ to be flat (uniform).
They are well-determined by the data and we  have no real prior
information that would override the data.

Failure to take the spatial distribution of the stars into account would result in the distances being
affected by the {\em Lutz-Kelker bias} (Lutz \& Kelker 1973)\nocite{lkbias}. But this can be accounted for
automatically by choosing the prior on $s$ based on our knowledge of the spatial distribution of the
population of stars (Smith 1999)\nocite{smi99}. The spatial distribution of Cepheid
variables is known to be flattened with respect to the Galactic plane. There are two
recent, observationally-supported, determinations of the exponential scale height $z_0$. 
Luri {\em et al.} (1999) \nocite{luri} obtained $z_0=97 \pm 7$ parsecs and
Groenewegen \& Oudmaijer (2000)\nocite{gw2000} determined $z_0=70 \pm 10$ parsecs.  We 
adopted $z_0=70 \pm10$ parsecs and introduced $z_0$ as an additional parameter. Our prior on
the distance looks like
\begin{displaymath}
p(s)\propto \rho(s)s^2ds,
\end{displaymath}
where $\rho(s)$ is the spatial density of stars:
\begin{displaymath}
\rho(s) \propto \exp\left( -|z|/z_0\right)
\end{displaymath}
with $z=s\sin\beta$, and $\beta$ is the galactic latitude of the
star.

The constant terms $u_0$ and $v_0$ get a flat prior. Unlike  the terms in
sines and cosines, which represent the physics of the pulsations, they are
just intercepts reflecting an arbitrary choice of coordinates. The
priors on the periodic Fourier coefficients $\mbox{\bf a}_u$ and
$\mbox{\bf a}_v$ must be chosen carefully. If our prior is too  vague,
significant terms may be rejected, but if it is too sharp, overfitting
may result. For our models we have used a Zellner G-prior, which is
equivalent to a maximum entropy prior (Gull 1988), \nocite{gul88} of the form
\begin{displaymath}
p(\mbox{\bf a}|\tau_a) \propto \tau_a^L\exp\left(-\frac{\mbox{\bf a}'\mbox{\bf
X}'\mbox{\bf Xa}}{2}\tau_a
\right)
\end{displaymath}
where $\mbox{\bf a}$ is the vector of Fourier coefficients, $\mbox{\bf X}$
is the design matrix of sines and cosines for the 
problem, $L=M$ or $N$ depending on whether the prior is
on the velocity or photometry coefficients, and
$\tau_a$ is another hyperparameter.

These hyperparameters $\tau_a$ (one for velocities, $\tau_a{_u}$, one for photometry, 
$\tau_a{_v}$) also need priors. Since they are scale parameters, one might naively put 
a $1/{\tau_a}$ Jeffreys prior on these; however, the resulting posterior distribution 
would be improper, due to a logarithmic divergence at $\tau_a=0$ (Gull 1988).\nocite{gul88} 
This is a common feature of hierarchical Bayes problems such as ours. To avoid this a 
slight adjustment is required. We  pick a prior on $\tau_a$ of the form
\begin{equation}
p(\tau_a) \propto \frac{1}{\tau_a^{3/2}}
\end{equation}
Strictly speaking there might still be a possibility of obtaining an improper posterior 
distribution due to the ``tail'' of this distribution,  so some researchers include an 
exponential damping term; however in our case this is not necessary.

\subsection{Sampling strategy}

Fortunately, the full conditional distributions for the precision parameters and the
hyperparameters are standard $\chi^2$ distributions and so the sampling for these parameters
can be accomplished with straightforward Gibbs sampling; that is, these parameters are updated
by draws from the respective, complete, conditional posterior distributions.

We use a random-walk Metropolis--Hastings algorithm to sample $\Delta\theta$, $\phi_0$ and $s$
simultaneously, using as our proposal distribution a multinormal distribution centered on the
currently imputed parameter values, with a variance-covariance matrix
that is proportional to the variance-covariance matrix for the linearized
least-squares problem for just these three parameters. This means
linearizing the logarithm in the expression  for $c_j$ (Eq.
\ref{eq:stefan}). The idea behind this strategy is that we'll take longer
steps in directions with larger variances and shorter steps in directions
with smaller variances, while obtaining good sampling in directions that
are not parallel to the axes defined by the parameters, so that if there
are significant correlations between these three variables we will still
sample them efficiently. This turns out to have very good
acceptance-rejection probabilities and good sampling of the parameter
space for these parameters. 

The sampling for $\mbox{\bf a}_u$ and $\mbox{\bf a}_v$ is more direct. We
base our proposal for a  Metropolis step on the solution of the linear
least squares problems generated by
\begin{eqnarray*}
& \mbox{\bf U} \sim \mbox{\bf
N}(u_0+\mbox{\bf X}_u\mbox{\bf a}_u,\sigma_U^2/\tau_U)\\  
& \mbox{\bf V} \sim \mbox{\bf
N}(v_0+\mbox{\bf X}_v\mbox{\bf a}_v,\sigma_V^2/\tau_V)  
\end{eqnarray*}
This results in a near-Gibbs sampler for these parameters. It isn't quite 
Gibbs because of the nonlinear way in which $\mbox{\bf a}_u$  and
$\mbox{\bf a}_v$ appear in the full likelihood. However, it is very
close; the acceptance probabilities for these proposals are over 90\%,
and the sampling of the Fourier parameter space is very effective. Again,
by sampling the parameters simultaneously, we will be able to sample
efficiently even if they have significant correlations.

Updating the Fourier coefficients we run into an additional complication. Updating the 
unknown order of M and N, respectively, of the Fourier approximation requires us to 
consider MCMC simulation across models of differing dimensional parameter space. For 
example, if we want to consider incrementing M, we need to add additional coefficients
in $\mbox{\bf a}_u$. MCMC schemes that allow such moves across different dimension models are
known as reversible jump MCMC (Green 1995;\nocite{green} Carlin \& Chib 1995).\nocite{cc1995}
See, for example, Dellaportas, Forster, \& Ntzoufras (2002) \nocite{del02} for a tutorial
review. We include a reversible jump MCMC move with the steps in $\mbox{\bf a}_u$ and
$\mbox{\bf a}_v$. Thus, if the current model has a certain number of parameters, we propose a
jump to a model with a (possibly different) number of parameters and simultaneously propose
new values for all the Fourier coefficients. To make the sampling efficient, during the
burn-in phase we also estimate the posterior probabilities of the individual models. We use
this as the basis for the proposal probabilities of new models during the computation phase
of the calculation. Thus we will propose models of higher posterior probability with greater
frequency.

\section{Cepheid data}

\subsection{Selection of Cepheids}

The thirteen Cepheids selected for our analyses were chosen because of their high
quality photometry and availability of radial velocities, although one of the
velocity curves is of poor quality (RS Pup). The latter permitted testing our
analysis with less than ideal data. In addition, we desired a large range in
pulsation period, a range in interstellar extinction, and inclusion of possible
overtone pulsators (SZ Tau, EU Tau).  The Cepheids selected and some quantities of interest
are listed in Table 1. These quantities and the individual stars are discussed in the following
subsections.

\subsection{Photometry and Radial Velocities}
For each Cepheid we require photometric magnitude and color data and radial
velocities.  These data were collected from the recent literature with much help from
the McMaster Cepheid Photometry and Radial Velocity Archive \\
(URL http://dogwood.physics.mcmaster.ca/Cepheid//HomePage.html) and from electronic 
transfers of data by N. Samus and L. Berdnikov.

We chose to use all photometric data for which HJD, $V$, and $(V-R)$ were available. Photometry
acquired in the Cousins $(V-R)_c$ system was converted to the Johnson $(V-R)_j$ system using
the precepts of Taylor (1986) \nocite{taylor86} as given in his Table 4:

\noindent
in right ascension zone 01:20--10:00 hours
\begin{equation}
\left(V-R\right)_j=\left(\left(V-R\right)_c+0.035\right)/0.714
\end{equation}
\noindent
in right ascension zone 15:30--18:00 hours
\begin{equation}
\left(V-R\right)_j=\left(\left(V-R\right)_c+0.023\right)/0.714 \label{eq:rysco}
\end{equation}
\noindent
for all other right ascension values
\begin{equation}
\left(V-R\right)_j=\left(\left(V-R\right)_c+0.029\right)/0.714
\end{equation}

\noindent
These relations shift the zero point of the $(V-R)_j$ scale by $\pm 0.008$ magnitude
for Cepheids in the two right ascension zones.  This has a negligible effect upon
our results.  Because all the data in our sample have been transformed, we will
hereafter drop the subscript $j$ on the Johnson $(V-R)$.

We used all radial velocity data in the literature published in 1980 and later. The factor,
$p$, that converts radial velocity to pulsational velocity was adopted from
Gieren, Barnes, \& Moffett (1989a)\nocite{gieren89a}.
\begin{equation} \label{eq:pvalue}
p=1.39-0.03\log P
\end{equation}
\noindent
where the period $P$ is in days.  This relation is an approximation to the theoretical
values for Cepheids derived by Hindsley \& Bell (1986)\nocite{hb86}.  The range in $p$ for
the Cepheids in this paper is 1.34--1.38, with the lower values of $p$ corresponding to the
longer pulsation periods.

We require an uncertainty for each value of $V$, $(V-R)$ and radial velocity, $V_r$.  The
uncertainties in the original sources were adopted when given, estimated when not given,
and equated to the previously defined parameters $\sigma_{V}$, $\sigma_{C}$ and
$\sigma_{U}$, respectively.

In most cases the photometry and velocities were not obtained simultaneously, requiring care
in the choice of pulsation period to phase the data properly. The periods listed in
Table 1 yielded light, color and velocity curves with no obvious systematic effects between
sources, unless otherwise noted.  

\placetable{1}

\subsection{Interstellar reddening}

$E(B-V)$ values were taken from Fernie {\em et al.} (1995) \nocite{fbes95} as tabulated at
URL\\ http://ddo.astro.utoronto.ca/cepheids.html.  Following the recommendation at that
site, we chose the values in column FE1 which are the $E(B-V)^{clus}$ values of Fernie (1990).
\nocite{f90} These reddenings are listed in Table 1.

We took the ratio of total--to--selective absorption, $R$, from the work of Gieren,
Barnes, \& Moffett (1993),\nocite{gbm93} which was itself adapted from Olson (1975), \nocite{o75}
\begin{equation}
R=3.15+0.25\left(<B>-<V>\right){\rm_0}+0.05E\left(B-V\right) \label{eq:R}
\end{equation}
\noindent
Values of $(<B>-<V>)$ were taken from the Fourier fits by Moffett \& Barnes
(1985). \nocite{mb1985}  The range of $R$ values for the Cepheids in our sample is 3.28--3.42 with mean
3.35.

With that range of $R$ the Cardelli, Clayton, \& Mathis (1989) \nocite{cardelli89} reddening
law gives
\begin{equation}
E(V-R)/E(B-V) =0.80{\pm 0.02}, 
\end{equation}
\noindent
reasonably close to the ratio 0.84 obtained earlier by Johnson (1968). \nocite{johnson}  We have
adopted the latter value for consistency with our earlier work.

\subsection{Individual Cepheids}

In Table 2 we tabulate both the number of observations and the sources of the photometry and
radial velocities.  In the following we note any peculiarities in the data that required
attention.  

\placetable{2}

T Mon --- We did not use the photometry of Berdnikov, Ignatova, \& Vozyakova (1998)
\nocite{be1998} because we could not satisfactorily transform their $(V-R)_c$ to the Johnson
system for this star.  (In general we were able to transform their photometry successfully; when we 
were not, as for T Mon, we note that in the discussion of the individual Cepheids.) We adopted the
parameterization of the pulsation period given by Evans {\em et al.} (1999); \nocite{evans99} {\em i.e.},
prior to
$HJD=2445700$ the period is 27.019389 days and subsequently, 27.032881 days. In our  tables and plots we
characterize the period as 27.026 days. 

T Mon is the only Cepheid in our sample with a binary orbit. Evans {\em et al.}
(1999) \nocite{evans99} give two acceptable orbits; one with an orbital period of 93987$\pm
17972$ days and another with 33650$\pm 381$ days. The orbital contribution to the radial
velocity is essentially the same for both orbits over the time span of our analysis.
We adopted the longer period orbit and corrected the observed velocities accordingly. 

RS Pup --- According to Szabados (1999, private communication) the period of RS Pup is highly
unstable.  We computed a period 41.467$\pm 0.002$ days from our photometric data set that
also fits the relatively poor radial velocity data. This may be compared to the period 41.386
days given by the GCVS1987 and 41.415 days listed by Moffett \& Barnes (1985).
\nocite{mb1985}

U Sgr --- U Sgr shows some evidence of binarity in the radial velocities; however, no orbital
solution has been found. Using Table 1 of Bersier {\em et al.} (1994) \nocite{bersier94} we
attempted a correction for the binary orbit, but this greatly degraded the velocity curve. 
We chose to ignore the binary correction.  

WZ Sgr ---  The color curve shows some modest systematic differences among the sources, but
we did not attempt to address this.  

BB Sgr --- The shape of the velocity curve differs slightly between Gieren
(1981a), \nocite{gieren1981a} on the one hand, and Gorynya {\em et al.} (1998),
\nocite{ge1998} Gorynya {\em et al.} (1996), \nocite{ge1996} and Lloyd Evans
(1980) \nocite{lloydevans1980} on the other. We did not attempt to reconcile the difference.

RY Sco --- The $(V-R)_c$ transformation had to be modified to map the Cousins data of
Berdnikov \& Turner (2000) \nocite{bt2000} and Coulson \& Caldwell (1985) \nocite{cc1985} to
the Johnson data of Moffett \& Barnes (1984). \nocite{mb1984} We changed Eq.(\ref{eq:rysco})
to 

\begin{equation}
\left(V-R\right)_j=\left(\left(V-R\right)_c+0.043\right)/0.714
\end{equation}

SZ Tau --- The photometric data in Berdnikov, Ignatova, \& Vozyakova (1998) \nocite{be1998}
and Berdnikov, Ignatova, \& Vozyakova (1997) \nocite{be1997} did not satisfactorily
transform from $(V-R)_c$ to $(V-R)_j$ and were not used. We computed a new period that fit
the photometric data well.  Our period of 3.14895$\pm 0.00002$ days differs by -0.00019 day
from that found for earlier epochs by Szabados (1991). \nocite{sz91}  Szabados noted
that SZ Tau shows erratic period variation larger than the change adopted here.  The new
photometric period does not fit the full velocity data set, which covers a longer time
interval than the photometry.  On the other hand, a best fit period to the velocities does
not adequately fit the photometry.  To resolve this, we applied our photometrically determined
period to both the photometry and the velocities.  Then velocities obtained after $HJD=2447500$
were shifted by +0.06 in phase to bring them into phase agreement with earlier velocities. 

EU Tau --- Photometric data in Berdnikov, Ignatova, \& Vozyakova (1998) \nocite{be1998}
and Berdnikov, Ignatova, \& Vozyakova (1997) \nocite{be1997} did not satisfactorily
transform from $(V-R)_c$ to $(V-R)_j$ and were not used. We adjusted the radial velocities of
Gieren {\em et al.} (1989b), \nocite{gieren89b} as suggested in their paper, by +3.05 km/s.  

T Vul ---  We adjusted the radial velocities of Bersier {\em et al.} (1994)
\nocite{bersier94} by +3 km/s to force agreement with the other velocity data. 

SV Vul --- The quadratic period given by Szabados (1991) \nocite{sz91} did not fit the
current data set. We computed a new ephemeris for the photometric and radial velocity
data used here that fits both reasonably. The variation we obtained is
\begin{equation}
HJD_{max}=2443716.383+45.0850E-0.000600E^2, 
\end{equation}
\noindent
which may be compared to that of Szabados (1991) \nocite{sz91}
\begin{equation}
HJD_{max}=2443716.383+45.0068E-0.000364E^2
\end{equation}

\noindent
In our tables and plots we characterize the period as 45.019 days. As Szabados (1991)
\nocite{sz91} has noted, the instantaneous period of SV Vul varies about the mean period by
several  days. This induces a scatter that is readily apparent in our photometric light and
color curves.   

\section{Results}

The calculations were run on a Macintosh G4 computer under system MacOS 9.1 using the R-1.4.1
language distributed by the R Development Core Team (See URL http://www.R-project.org/).  
(R-1.4.1 is available for several platforms.) For all stars we adopted a ``burn-in'' of 1000 
samples and a  total number of 10,000 samples.  Output quantities and their uncertainties are shown
in Tables 3, 4 and 7. In this section we discuss these quantities.

\subsection{Analysis diagnostics}

In Table 3 we list diagnostic parameters, {\em i.e.} quantities that inform us about the
quality of the data and of the analysis. The quantities given are the phase difference between
the photometry and velocities ($\Delta\theta$) and its uncertainty, the most probable number
of terms in the $V$ magnitude Fourier model ($N$), the most probable number of terms in the velocity
Fourier model ($M$), and the square roots of the hyperparameters ($\sqrt{\tau}$) on the photometric and
velocity uncertainties.  We will discuss these in turn.

\placetable{3}

\subsubsection{Phase shifts}

Recall that our model permits a difference in phase ($\Delta\theta$) between the
Fourier series model for the photometry and that for the velocity.  (The offset $\Delta\theta$ is 
{\em added} to the displacement curve phases, thus it is an adjustment to the phases of the
velocities relative to the photometry.)  $\Delta\theta$ allows for the possibility that we
have adopted an incorrect period in phasing the data or that the period has changed between the 
epoch of the photometry and that of the velocities. It is obvious that
an incorrect period, coupled with photometry and velocities acquired at different epochs, would lead to a phase
shift between the modeled photometry and the modeled velocities. However,
Barnes {\em et al.} (1977) \nocite{bde1977} have shown that an incorrect slope in the surface
brightness relation, Eq. \ref{eq:nordgren}, can also lead to a non-zero $\Delta\theta$.  It
is important that we determine whether either of these possibilities is true.  

Seven of the thirteen Cepheids show a non-zero $\Delta\theta$, with estimated posterior mean more
than two posterior standard deviations away from zero. We will first enquire whether these seven may
be the result of period errors.  If the adopted period is incorrect or has changed, we expect a
significant $\Delta\theta$ value to arise when the photometry  and the velocities are not coeval.  To
test this we computed for each of the seven Cepheids the mean date of the photometric observations,
the mean date of the velocity observations, and the change in period required to account for the
$\Delta\theta$ listed in Table 3 over that interval.  For five of the stars (excluding X Cyg and BF
Oph) the change in period required to account for $\Delta\theta$ is comparable to the uncertainty in
the period. In most of these cases the fault is not so much a poor period, as a large interval
between the photometry and the velocities given the period uncertainty.  For these five we attribute
the apparently significant $\Delta\theta$ to uncertainty in the period coupled with a large interval
between the photometric observations and the velocity observations.  

For X Cyg and BF Oph, the required change in period is seemingly much larger than the
uncertainty in the period will support.  In the case of X Cyg, Evans (1984)
\nocite{evans1984} has shown that systematic phase shifts as large as 0.02 period occur. While 
the origin of these phase shifts is unknown, they may account for the $-0.021$ phase shift we 
find over the 85 cycles (nearly 4 years) that separate the photometry and the velocities.  For BF Oph the
period analysis by Szabados (1989) \nocite{sz89} showed some systematic effects in the residuals, but
they are much smaller than the $\Delta\theta$ found in our analysis.  An explanation for the
large $\Delta\theta$ for BF Oph that is based on an incorrect period is not supported by Szabados'
analysis.

Thus we find for twelve of the thirteen Cepheids $\Delta\theta$ values that are zero within their
uncertainties or have significantly non-zero values that may be attributed to a large gap
between the photometric observations and the velocity observations.  The phase shift for BF
Oph is unexplained. We conclude that the $\Delta\theta$ values found here give us no reason to
believe that the adopted slope in the surface brightness relation, Eq. \ref{eq:nordgren}, is
incorrect.  

Nonetheless, this is such an important issue that we will enquire whether the mean
$\Delta\theta$ is consistent with the published  uncertainty in the slope of the surface
brightness relation. The variance weighted mean $\Delta\theta$ for all thirteen stars is
$-0.0115\pm0.0065$(s.e.m). Were we to discard the six stars with $\Delta\theta$ values likely
affected by non-coeval photometry and velocities, the mean would be $-0.0034\pm0.0099$.  A
simulation published by Barnes {\em et al.} (1977) \nocite{bde1977} suggests that
$\Delta\theta$ of order $-0.01$ would be caused by a change in the slope of the surface
brightness relation of order 0.006---0.010. We ran simulations on all our stars and found that
a phase difference of $-0.011$ results from a change in the slope of $-0.012$. With a mean phase shift in
the range $-0.0034$ to $-0.0115$, the implied error in the slope in Eq. \ref{eq:nordgren} is $-0.004$ to
$-0.013$ with the former value being more likely because of changing or erroneous periods.  Recall that the
uncertainty in the slope of the surface brightness relation, Eq.\ref{eq:nordgren}, is $\pm 0.007$, as given
by Nordgren {\em et al.} (2002)\nocite{nordgren2002}.   We conclude that the phase shifts found here are
consistent, within the error, with the adopted slope in Eq.\ref{eq:nordgren}.

\subsubsection{Fourier Polynomial Orders}

As discussed in Section 4.3 we modeled the $V$ light curve and the radial velocity curve of
each Cepheid as Fourier polynomials of order $N$ and $M$, respectively. In about half the cases
only one polynomial order had a significant posterior probability and that order is listed in
Table 3.   When multiple orders are listed, the posterior probability is above 5\% for all
those orders, and they are listed in descending order of the posterior probability. In this 
section we demonstrate that our analysis reasonably models these  curves.  

We first demonstrate that our Bayesian analysis chooses a model-based Fourier order that agrees with
the order that a researcher would have chosen from exploratory data analysis of the Fourier
polynomial against the data.  One way to demonstrate this is to compare the orders selected in our
analysis with the orders chosen by the traditional method.  In the traditional analysis the order is
selected based on a visual impression of the fit to the data, usually with the criterion that the
polynomial be terminated when the scatter about the polynomial approaches the uncertainty in the
data. Note that, because the light and velocity curves are in reality not finite Fourier series,
fitting until the noise level is reached may overfit the curves and lead to excessively large values of
$N$ and $M$. 

Moffett \& Barnes (1985)\nocite{mb1985} have published Fourier series coefficients for the light curves
for twelve of our thirteen stars.  A Fourier fit for the light curve of the thirteenth star, EU Tau, has
been published by Gieren {\em et al.} (1990). \nocite{gmbfm1990} In Figure \ref{fig1} we plot the order of
the $V$ magnitude Fourier polynomial determined here against the order determined in the works just
cited. In some cases our results give a significant posterior probability for two values of $N$; in those
case we have plotted a fractional $N$ determined by weighting by the posterior probabilities. The
agreement is excellent, especially when one considers the additional photometry included in our analyses
that was not available to the earlier researchers and the possibility of overfitting in previous work. 
We conclude that the Bayesian analysis successfully and objectively chooses the order of the Fourier
polynomial for the $V$ magnitude data and therefore for the velocity data also.

Another way to demonstrate the success of our objective order selection is to plot the Fourier polynomial
for that order with maximum posterior probability against the data and perform the customary `eye ball'
test.  The posterior probabilities for the photometry and velocity models for X Cyg are shown in Figures
\ref{fig2} and \ref{fig3}.  For the photometry both the ninth and tenth order models are supported by the
data.  For the velocities, eighth, ninth and tenth order models are supported. In Figures \ref{fig4} and 
\ref{fig5} we show comparisons of the data with Fourier polynomials of order nine.  The comparisons 
are quite reasonable, given that there is significant posterior probability for models with order ten
(photometry) and orders eight and ten (velocities).

\subsubsection{Hyperparameters $\tau$}

Earlier we introduced hyperparameters on the photometric and velocity uncertainties, {\em
i.e.}, we chose not to trust the uncertainties published or adopted for the individual data, and we
modeled the extent to which the uncertainties describe the actual scatter in the
observations. The hyperparameters $\tau$ are defined such that the posterior variances in velocity,
in $V$ magnitude, and in $(V-R)$ color data are given respectively by $\sigma_{U_i}^2/\tau_U$,
$\sigma_{V_j}^2/\tau_V$, $\sigma_{C_j}^2/\tau_C$ where the $\sigma$ values are the published or adopted 
uncertainties.  Therefore $\tau$ $\le 1$ implies that the actual variance of the data about
the Fourier polynomial fit is greater than the variance given by the adopted uncertainties.  

As shown in Table 3, this is overwhelmingly the case. There we show the mode of each posterior
probability distribution for the hyperparameters. (In Table 3 we actually quote $\sqrt{\tau}$ as it
is more closely related to the published standard errors.)  Either the adopted uncertainties
underestimate the actual scatter in the data, or there is additional scatter in our photometric and
velocity curves not represented by the observational errors.  (To some extent this may be caused by
model misspecification, {\em i.e.}, we know that these pulsation curves are not finite Fourier series.) 
Additional scatter could be induced by period jitter affecting the pulsation phases, by errors in
the  transformation from Cousins-to-Johnson photometry, by mismatches between the photometric and
velocity systems of the original sources, {\em etc}. 

Period jitter is clearly seen in our magnitude and velocity curves for SV Vul. This is a 
result of unexplained, short time scale variation in the pulsation period, as noted in Section
5.4. It is the likely cause of SV Vul having the smallest $\tau_U$ and the next to smallest
$\tau_V$. No other star shows evidence of period jitter in examination of the photometry and
velocities.

Were the dominant contributor to the excess scatter the Cousins-to-Johnson transformation, the
largest effect would be seen in $(V-R)$, as only the color data are affected by the transformation.  Just
the opposite is seen; $\sqrt{\tau_C}$ is generally larger than the other two.  We conclude that our color
transformation is not a significant source of additional scatter. 

Examination of the individual stellar values of $\sigma_V$ and $\sigma_C$ reveals that the claimed
observational uncertainties are quite small.  For example, the mean values for 330 magnitude and color
observations of  U Sgr are $\pm 0.016$ mag. and $\pm 0.018$ mag., respectively.  In our experience it is
quite difficult to achieve such  precision in the Johnson magnitudes, especially for large-scale
observational programs such as those from which we drew the data for this paper. Furthermore, it is well
known that photometric magnitudes are less precise than photometric colors, when obtained with
photoelectric photometers, due to cancellation of some error sources in determining the color.  That is
not borne out in most of these data.  Examination of the values of $\sqrt{\tau_C}$ and $\sqrt{\tau_V}$
shows that the color hyperparameter is closer to unity than the magnitude hyperparameter in twelve of
thirteen cases. These facts suggest that a reasonable interpretation of the $\tau$ values is that the
published or adopted uncertainties in the photometry are underestimated.  Adopting this interpretation
and the values in Table 3, we estimate that the true observational scatter for U Sgr is typically
$\sigma_V/\sqrt{\tau_V} = \pm 0.030$ mag. and $\sigma_C/\sqrt{\tau_C} = \pm 0.024$ mag. These are
reasonable values for observational programs of the type listed in Table 2. While we cannot rule out all
possible sources of excess scatter in the photometric and velocity curves, we believe that underestimated
observational uncertainties are the most likely explanation for $\sqrt{\tau}$
${\le 1}$.  

\subsection{Radius results}

In this section we present our results on the radii and angular diameters.  Table 4 lists the
mean stellar radius ($<R>$) in solar radii and the mean stellar angular diameter ($<\phi>$) in
milliarcseconds, and their uncertainties. Angular diameters  are included here for the
benefit of researchers making direct measurement of Cepheid angular diameters.  The mean uncertainty 
in a radius is $\pm 7.7\%$, and without the two outliers RS Pup and SZ Tau, it reduces to $\pm 6.5\%$. 

\placetable{4}

A least-squares fit to our results, weighted by the variance in $log(<R>)$ and excluding SZ
Tau and EU Tau as possible overtone pulsators, gives the fundamental-mode period-radius (PR)
relation from 11 Cepheids
\begin{equation}
log(<R>) = 0.679{(\pm 0.050)}(log(P)-1.2)+2.044{(\pm 0.065)},  
\end{equation}
\noindent
We have placed the zero point of the PR relation at $log P = 1.2$, at the
mid-point of our distribution of periods. Alcock  {\em et al.} (1995) \nocite{alcock} have
determined the ratio between first overtone and fundamental period for Galactic Cepheids,
which is $0.702$ and $0.707$ for SZ Tau and EU Tau, respectively.  If we include SZ Tau and EU
Tau in the PR relation with their equivalent fundamental-mode periods, we obtain for 13 Cepheids 
\begin{equation}
log(<R>) = 0.693{(\pm 0.037)}(log(P)-1.2)+2.042{(\pm 0.047)},  
\end{equation}
\noindent
In anticipation of the discussion below of the pulsation status of SZ Tau and EU Tau, we will
adopt this as our best estimate for the PR relation and show it in Figure \ref{fig6}.

The weighted rms scatter in Figure \ref{fig6} is $\sigma=0.040$, compared to the typical scatter in the
$log<R>$ values of $\sigma=0.033$.  This implies an additional scatter in the PR relation of about
$\pm0.02$.  Excess scatter of this order or more is typically found in Cepheid PR relations (see
citations in Table 5).

\placetable{5}

In Table 5 we collect recent determinations of the Cepheid PR relation. We have adjusted these
relations to our definition of the factor $p$ given in Eq. \ref{eq:pvalue}.  We have also recomputed the
uncertainties in some cases. The weighted mean PR relation from Table 5 is 
\begin{equation}
log(<R>) = 0.690{(\pm 0.018)}(log(P)-1.2)+1.979{(\pm 0.006)},\label{eq:PR}
\end{equation}
\noindent
(In assigning a weight to the result from theoretical models, we set the uncertainty to a median value 
so that solution would not dominate the mean.) 

It is apparent from the consistent values of the zero point in Table 5 that all methods yield the same
radius for a Cepheid of period near 16 days ($log P = 1.2$).  With regard to the slopes, however, there
is much variation in the results.  It is beyond the scope of this paper to resolve that question, but it 
warrants further investigation.

Bono {\em et al.} (2001) \nocite{bono2001} have computed theoretical PR relations for first overtone
pulsators; with, and without, convective overshoot. Table 6 compares our observed radii (Table
4) for EU Tau and SZ Tau with those predictions and with the fundamental-mode PR
relation (Eq. \ref{eq:PR}).  Our radii are more consistent with overtone pulsation than with
fundamental-mode pulsation for both stars. For EU Tau our observed radius supports first overtone
pulsation without convective overshoot.  While our radius for SZ Tau agrees with that predicted for
overtone pulsation without convective overshoot, the larger uncertainty on its radius precludes
exclusion of the convective overshoot radius. 

\placetable{6}

\subsection{Distance results}

In this section we present our results on distance and parallax.  Table 7 lists
the stellar distance ($s$) in parsecs, the stellar parallax ($\pi$) in milliarcseconds,
the magnitude-mean absolute magnitude ($<M_V(mag)>$), and the intensity-mean absolute
magnitude ($<M_V(int)>$), and their uncertainties. The mean uncertainty in distance is $\pm
7.7\%$, and without the two outliers RS Pup and SZ Tau, it reduces to $\pm 6.3\%$.  These percentage
uncertainties are essentially the same as those in the radii, as expected.  The mean
uncertainty in absolute magnitude is $\pm 0.165$ mag. in accord with the mean distance
uncertainty.  

\placetable{7}

We illustrate the posterior probability distributions for the parallax and distance of \\ T Mon in Figures
\ref{fig7} and \ref{fig8}. It is apparent that the parallax probability distribution is symmetric and that
the distance probability distribution is skew.  This is to be expected as the two quantities are related
through the reciprocal. This means that our results in Table 7 for distance and parallax cannot be related
exactly through the reciprocal as each is drawn from its own posterior probability distribution. Our quoted
values are the expectation values from those distributions.  The skewed nature of the posterior
probability distribution in distance will not be found in analyses based on standard
maximum-likelihood techniques and shows an advantage of our Bayesian approach.

Figure \ref{fig9} shows the  simulation history of the parallax, demonstrating good acceptance rates for the
proposals. There is no evidence of `sticking' or of systematics in the sampling.  Figure \ref{fig9}
demonstrates that our proposals have been sampled efficiently and completely. Simulation histories on the
other parameters are quite similar.   

Figure \ref{fig10} shows the period-luminosity (PL) relation for our $<M_v(int)>$ results.
A solution weighted by the variance in $<M_v(int)>$ yields, for the eleven fundamental-mode Cepheids, 
\begin{equation}
<M_v(int)> = -2.682{(\pm 0.215)}(log(P)-1.2)-4.699{(\pm 0.280)} \label{eq:PL1}
\end{equation}
\noindent
If we include EU Tau and SZ Tau in the calculation with their equivalent fundamental-mode periods,
we obtain for 13 Cepheids
\begin{equation}
<M_v(int)> = -2.690{(\pm 0.169)}(log(P)-1.2)-4.699{(\pm 0.216)}  \label{eq:PL2}
\end{equation}
\noindent
We adopt this as our best estimate for the PL relation and show it in Figure \ref{fig10}. 

The weighted rms scatter in Figure \ref{fig10} is $\sigma=0.186$ mag. which may be compared to typical
uncertainty in Table 7, $\sigma=0.165$ mag. This indicates an additional scatter in the PL relation
of order $\pm0.09$ mag. beyond that present in our distance determinations. We attribute the additional
scatter to the presence of errors in the adopted interstellar extinction corrections and to the
fact that the PL relation is an approximation to a period-luminosity-color relation and thus has a
cosmic width.

In Table 8 we collect recent determinations of the Cepheid PL relation. Where appropriate we have 
adjusted these relations to our definition of the factor $p$ given in Eq. 19.  (The relation attributed 
to Turner \& Burke (2002)\nocite{tb2002} has been computed by us from the 23 $(M_v)_{BW}$ values in their
Table 3.)  The weighted mean PL relation from Table 8 is 
\begin{equation}
<M_v(int)> = -2.851{(\pm 0.056)}(log(P)-1.2)-4.812{(\pm 0.058)} \label{eq:PL}
\end{equation}
\noindent
Eqs. \ref{eq:PL2} and \ref{eq:PL} may be compared to the PL relation adopted by Freedman {\em et al.}
(2001)\nocite{freedman} for the final results of the HST key project on the extragalactic distance scale:
\begin{equation}
<M_v> = -2.760{(\pm 0.030)}(log(P)-1.2)-4.770{(\pm 0.030)}
\end{equation}

It is important to compare the predictions of these relations with the absolute magnitude inferred from
the Hubble Space Telescope parallax of $\delta$ Cep (Benedict {\em et al.} 2002)\nocite{benedict} and with
the absolute magnitude computed from the distance to $\delta$ Cep determined by Nordgren {\em et al.}
(2002).\nocite{nordgren2002} (Interferometrically-based distances to $\eta$ Aql and $\zeta$ Gem have
also been published, {\em viz.} Lane {\em et al.} 2002) {\nocite{lcn2002} We have adjusted the Nordgren
{\em et al.} distance to our adopted value of $p$. The comparisons are listed in Table 9. Given that
these values are based on different techniques with different interstellar extinction assumptions, it is
remarkable how well they agree with the HST parallax.  In particular, our value agrees nicely with the
HST value.  

\subsection{Astrophysical diagnostics}

In this section we examine the effects of several astrophysical choices on our results.  We test the
choice of Galactic scale height for the Cepheid distribution in our  prior on distance, the effect of
interstellar reddening errors on the results, and the effect of permitting a phase shift between the
velocities and photometry.

As discussed in Section 4.4 we chose a prior on the distance that depends on the scale height 
of the Cepheid distribution. The two observationally supported choices were $z_0=97 \pm 7$ parsecs 
obtained by Luri {\em et al.} (1999) \nocite{luri} and  $z_0=70 \pm 10$ parsecs from
Groenewegen \&  Oudmaijer (2000)\nocite{gw2000}. We selected the latter value.  We tested this choice
by running the analysis on SZ Tau with the alternate value of $z_0$. SZ Tau has $distance * sin(\beta) =
-180$ parsecs, the largest $|z|$ in our sample.  The result is that distance, radius, and $\Delta\theta$ 
change by $-3$ parsecs ($.03\sigma$), $-0.2$ solar radii ($.03\sigma$), and $-0.0027$ ($.12\sigma$)
pulsational period, respectively, which are entirely negligible compared to the uncertainties.  The
specific choice of Galactic scale height, within reasonable bounds, does not affect our results. However,
including it is crucially important as discussed in section 4.4.

One of the advantages of the visual surface brightness method is its insensitivity to errors in the 
adopted interstellar reddening.  This is easily demonstrated through Eq. \ref{eq:combined} with $A,B$ set
to their values in Eq. \ref{eq:nordgren} and with the reddening explicitly included in the $V_0$ and
$(V-R)_0$ terms. The prediction is that an error of $+0.10$ mag. in the value of $E(B-V)$ leads to an error
of $+1.1\%$ in the angular diameter. The errors in mean radius and distance are related to the error in 
angular diameter by 
\begin{equation}
{\delta\phi}/\phi = {\delta <R>}/<R> - {\delta s}/s
\label{eq:delta}
\end{equation}
\noindent
We tested the prediction by running our analysis on RY Sco with an assumed $E(B-V) = 0.877$ rather than
the actual value of $E(B-V) = 0.777$.  The resulting angular diameter, radius, and distance change by
$0.4\%$, $0.6\%$, and $0.2\%$, respectively.  While the angular diameter change is somewhat smaller than
the prediction, all three are  satisfactorily small and are related in accord with Eq. \ref{eq:delta}. For
comparison, note that an error in $E(B-V)$ of $+0.10$ mag. leads to an error in the distance determined
through the Cepheid PL relation of $-17\%$. Of course absolute magnitudes determined from visual
surface brightness distances contain the full effect of any errors in the reddening.

Finally, we examine the effect of $\Delta\theta$ on our results.  Use of such a phase shift permits
correction for erroneous or changing periods as we discussed earlier.  It has been suggested that a
non-zero $\Delta\theta$ may bias the distance results (Laney 1999, private communication).  It is important
to ascertain whether this occurs. To determine the extent to which our results depend on use of a variable
$\Delta\theta$, we modified the analysis to adopt phase agreement between the photometric and velocity
data and performed the analysis again on our sample of Cepheids. In Figure \ref{fig11} we plot the new
distances against the distances with $\Delta\theta$ allowed to vary (from Table 7).  The distances with 
$\Delta\theta=0$ are found to be $0.4\%\pm0.6\%$ larger in the mean than those with
$\Delta\theta=variable$.  It is clear that the use of variable $\Delta\theta$ has not materially affected
the distance results. The test had a similar result for effects upon the mean radius.

\subsection{Discussion and Conclusions}

We have constructed a fully Bayesian statistical approach to the determination of Cepheid variable
star properties through the surface brightness method.  This approach obtains a mathematically rigorous
solution to the problem, a significant improvement on previous attempts.  We have shown how priors that
represent our knowledge before looking at the data may be chosen and how the likelihood for the
available data may be constructed.  The posterior is then the product of the prior distribution and the
likelihood. Our analysis makes use of Markov Chain Monte Carlo simulation to generate posterior Monte
Carlo samples that can be used to evaluate any desired posterior integral.  We have presented an
implementation of Gibbs sampling and of Metropolis-Hastings sampling of the posterior distribution so
that the distribution may be sampled efficiently (shorter computing times) and completely (reliable
inferences). 

To demonstrate the Bayesian approach, we applied it to thirteen Cepheids for which appropriate data
exist in the literature.  The data required for our calculations are Johnson $VR$ photometry, radial
velocities, interstellar extinction estimates, and stellar pulsation periods.  These were drawn from the
literature, although we adjusted the pulsation periods for several stars.  

We chose to use the visual surface brightness relation in this test, although our analytic approach may be
used on other surface brightness relations, {\em e.g.} the infrared relation.  The relation adopted is
one recently calibrated by Nordgren {\em et al.} (2002) through interferometrically determined angular
diameters of several Cepheids.  They applied this calibration to the determination of the distance to
$\delta$ Cep, $272\pm6$ pc. The accuracy of their distance scale has been tested by Benedict {\em et al.}
(2002), who determined the distance to $\delta$ Cep $273\pm9$ pc through a Hubble Space Telescope parallax.
The $1\%\pm4\%$ agreement between these distance scales gives us considerable confidence in the Nordgren
{\em et al.} surface brightness relation. (When Nordgren {\em et al.}'s distance is adjusted to our adopted
value of $p$, the agreement is $4\%\pm4\%$.) Distances determined in our work with this visual surface
brightness relation are therefore very close to, in not on, a geometric distance scale.  

Built into our Bayesian approach are numerous opportunities to check the analysis, both analytical and
astrophysical checks. We discussed several of these: the phase difference between the photometry and the
velocities, the orders of the Fourier series used to model the photometry and the velocities,
hyperparameters on the observational uncertainties, the choice of scale height for the Cepheid distance
distribution, the effect of interstellar reddening on our results, and the effect of permitting a phase
shift on the distances and radii obtained.  In the next few paragraphs we describe these checks on our
results.

As part of our analysis, we determined phase shifts that optimally match the photometric and the radial
velocities curves.  All of these were found to be small, but about half were significant compared to their
uncertainties.  In six of the seven significant cases we show the $\Delta\theta$ value to be a result of
period uncertainty coupled with a significant interval between acquisition of the photometry and the
velocities.  Because an erroneous slope in the surface brightness relation can create phase shifts similar
to those we found, we examined that issue as well.  We showed that our phase shifts, if interpreted this
way, imply a change in the slope of the surface brightness relation less than its uncertainty.

We also showed that our model-based orders of the Fourier polynomial fits to the light curve
($N$) and the radial velocity curve ($M$) agree with previously determined (subjectively chosen)
values and that they produce fits to the curves that agree with our (subjective) judgement. The
ability to determine $N,M$ objectively is an important feature of our Bayesian approach, as is the
ability to average over models with different $N$ or $M$, each with significant posterior
probability.

Because we were unsure that the published observational uncertainties are correct, we introduced
hyperparameters into our model to check them.  The hyperparameters show clearly that there is more scatter
in the light, color, and velocity curves than can be explained by the published observational
uncertainties.   We examined several possibilities and concluded that the observational uncertainties are
likely to be underestimated. 

Our analysis takes into account the spatial distribution of Cepheids in a natural way through the choice of
distance prior.  This means that our distances and absolute magnitudes are not subject to Lutz-Kelter
bias.  We modified our choice for the scale height of the Cepheid Galactic distribution to inquire
whether the specific choice affects our results.  We found that a change from $70\pm10$ pc to $97\pm7$
pc, two recent determinations, has a negligible affect upon our distances and radii. 

A key advantage of surface brightness methods is their near independence from errors in the interstellar
extinction.  We showed that an error of $0.10$ mag. in $E(B-V)$ leads to errors in distance, angular
diameter, and radius of $+0.2\%$, $+0.4\%$, and $+0.6\%$, respectively. On the other hand, the same error
leads to a change in the distance determined through the Cepheid PL relation of $-17\%$. 

The final check we presented was to determine whether our use of a phase shift between the photometry and
the radial velocities had an affect on the distances and radii obtained. We performed all our calculations
afresh with $\Delta\theta$ fixed at zero. The distances with $\Delta\theta=0$ were found to be
$0.4\%\pm0.6\%$ larger in the mean than those with $\Delta\theta=variable$, which is entirely negligible. 
The radii showed a similar insensitivity.

From our sample of thirteen Cepheids we determined period-radius and period-luminosity relations.  The
radii we found for EU Tau and SZ Tau show strong support for interpreting them as overtone pulsators,
and, in the case of EU Tau, strong support for theoretical models without convective overshoot.  (The
radius of SZ Tau agrees with prediction from models without convective overshoot, but its uncertainty
precludes a definitive statement.) From all thirteen Cepheids we determined PR and PL relations
\begin{equation}
log(<R>) = 0.693{(\pm 0.037)}(log(P)-1.2)+2.042{(\pm 0.047)},  
\end{equation}
\noindent
\begin{equation}
<M_v(int)> = -2.690{(\pm 0.169)}(log(P)-1.2)-4.699{(\pm 0.216)} 
\end{equation}
\noindent
As a check on these results we compared them to consensus relations from the recent literature, {\em
ie},
\begin{equation}
log(<R>) = 0.690{(\pm 0.018)}(log(P)-1.2)+1.979{(\pm 0.006)},
\end{equation}
\noindent
\begin{equation}
<M_v(int)> = -2.851{(\pm 0.056)}(log(P)-1.2)-4.812{(\pm 0.058)}
\end{equation}
\noindent
An additional check is to compare our predicted absolute magnitude for $\delta$ Cep to that obtained by
Benedict {\em et al.} (2002) from an HST parallax. Our PL relation predicts $-3.43\pm0.10$ mag compared
to $-3.47\pm0.10$ mag. given by Benedict {\em et al.}.

We conclude that our Bayesian analysis is a powerful tool for determining properties and distances for
Cepheid variables through the surface brightness technique.

\acknowledgments
We thank L. Szabados for assistance in period selection for these stars. We thank 
Pawel Moskalik for pointing us to several useful radial velocity papers.
Electronic data transfers were generously provided by N. Samos and L. Berdnikov and
were greatly appreciated. Binary velocity corrections for T Mon were made possible by
computer code provided by Jos Tomkin, whom we thank for his help. This research made
use of the McMaster Cepheid Photometry and Radial Velocity Archive at
http://www.physics.mcmaster.ca/Cepheid/ which is maintained by D.L. Welch and is
supported, in part, by the Natural Sciences and Engineering Research Council of
Canada  (NSERC).

\clearpage


\begin{figure}
\figurenum{1}
\plotone{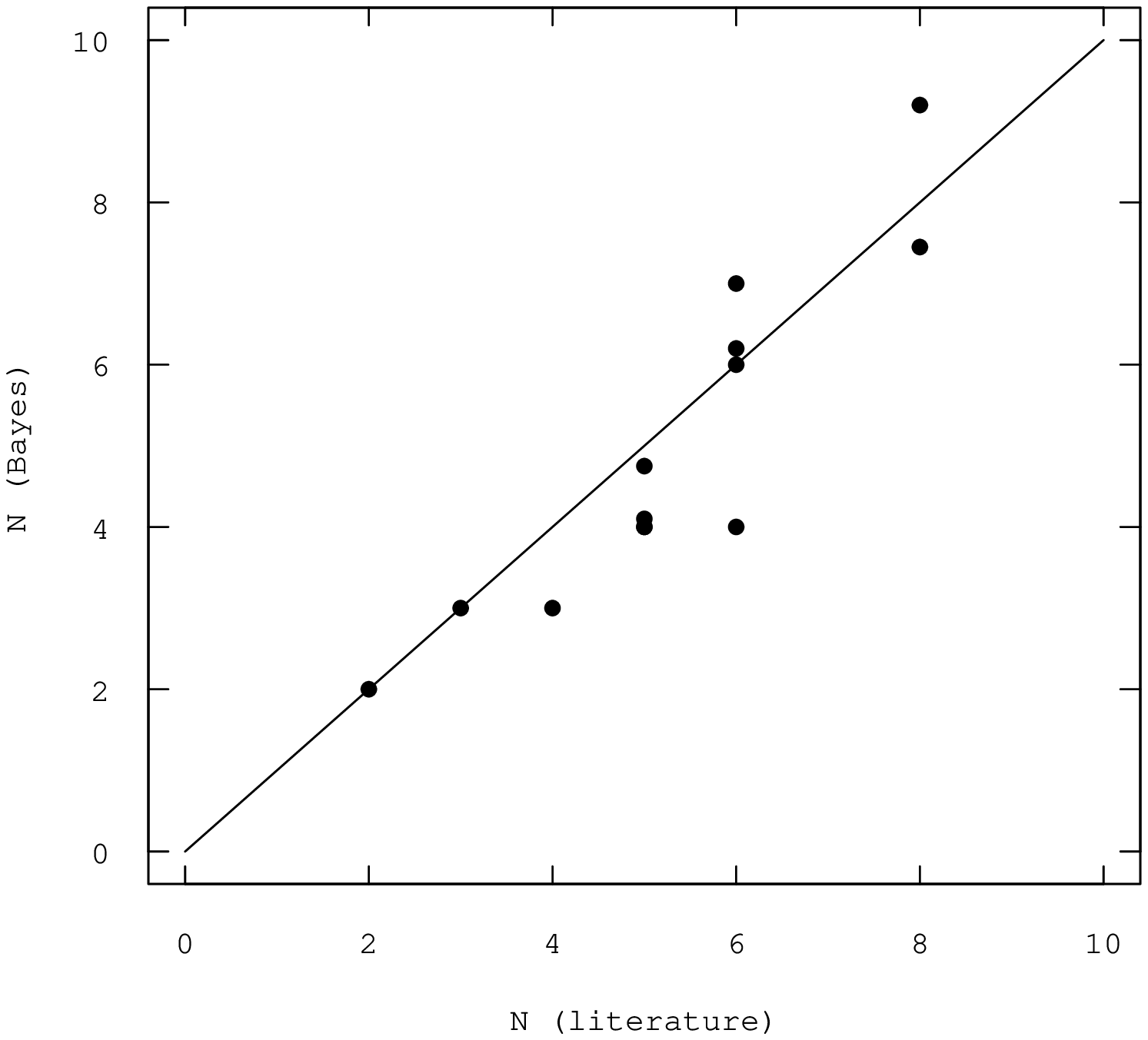}
\caption{Comparison of $V$ light curve Fourier orders from our Bayesian analyses with those from 
previous work. A line of equality is shown for illustration. \label{fig1}}
\end{figure}

\begin{figure}
\figurenum{2}
\plotone{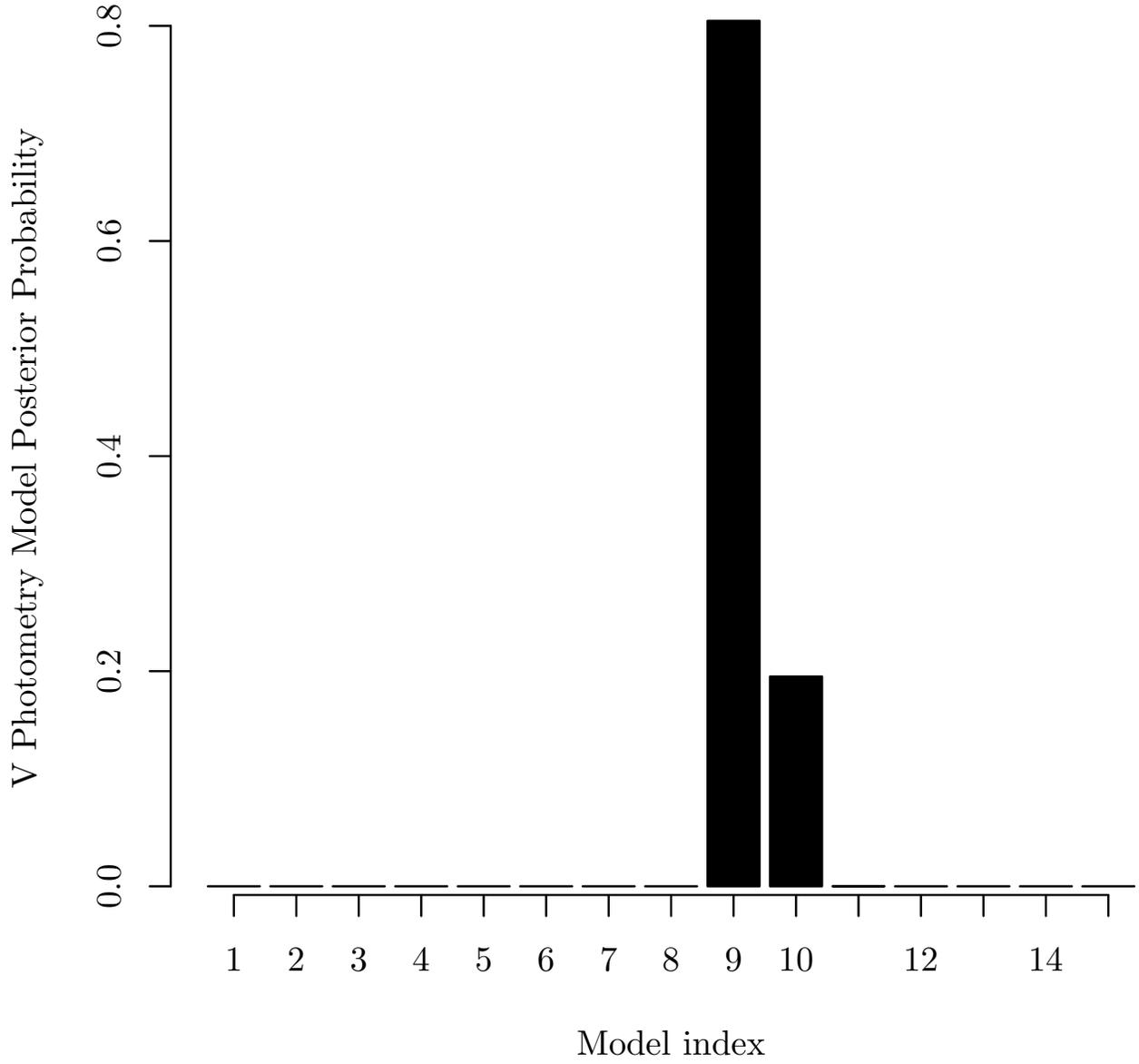}
\caption{Posterior marginal distribution of photometry models for X Cyg. \label{fig2}}
\end{figure}

\begin{figure}
\figurenum{3}
\epsscale{.8}
\plotone{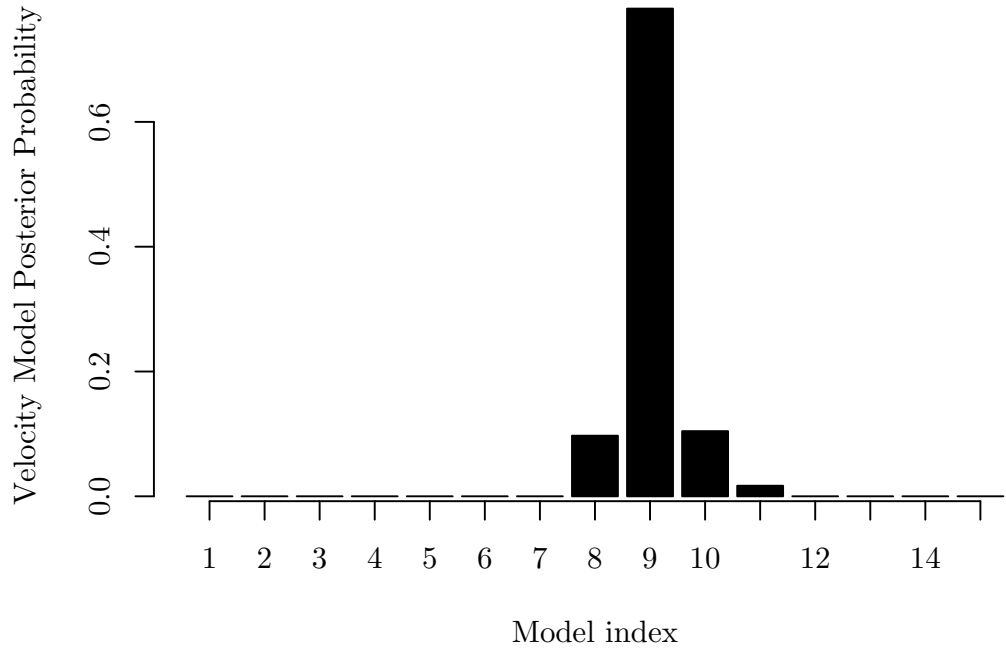}
\caption{Posterior marginal distribution of velocity models for X Cyg. \label{fig3}}
\end{figure}

\begin{figure}
\figurenum{4}
\plotone{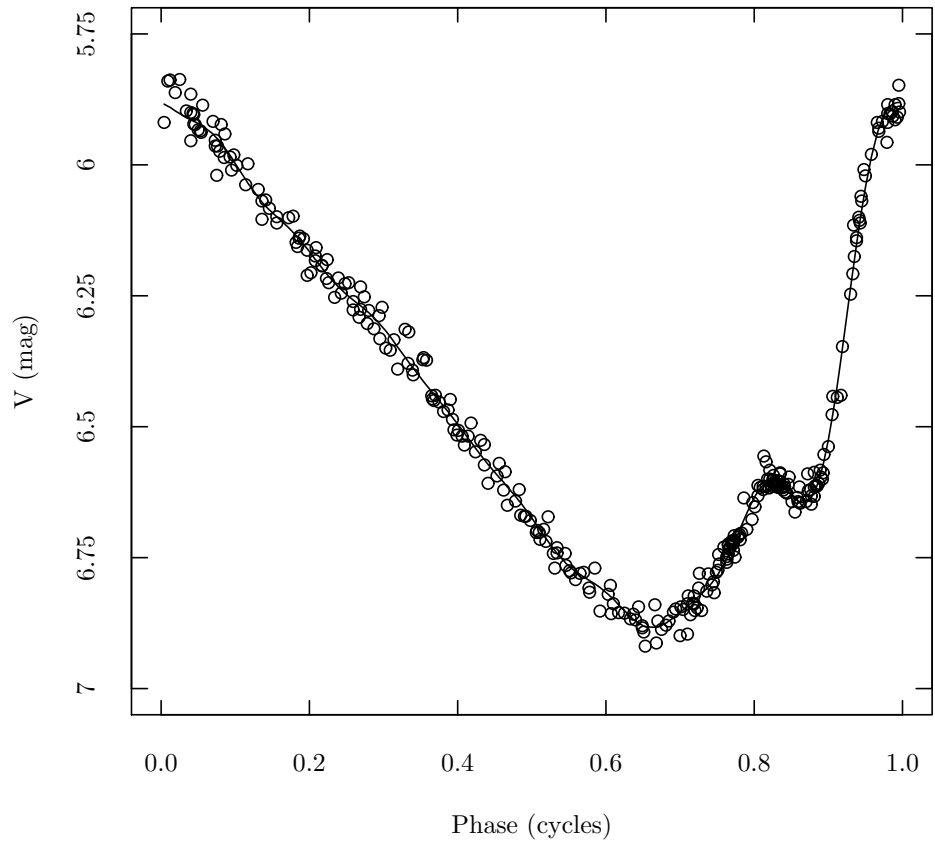}
\caption{The $V$ light curve for X Cyg.  A Fourier polynomial of order $N=9$ is shown fitted to the
observations.
\label{fig4}}
\end{figure}

\begin{figure}
\figurenum{5}
\plotone{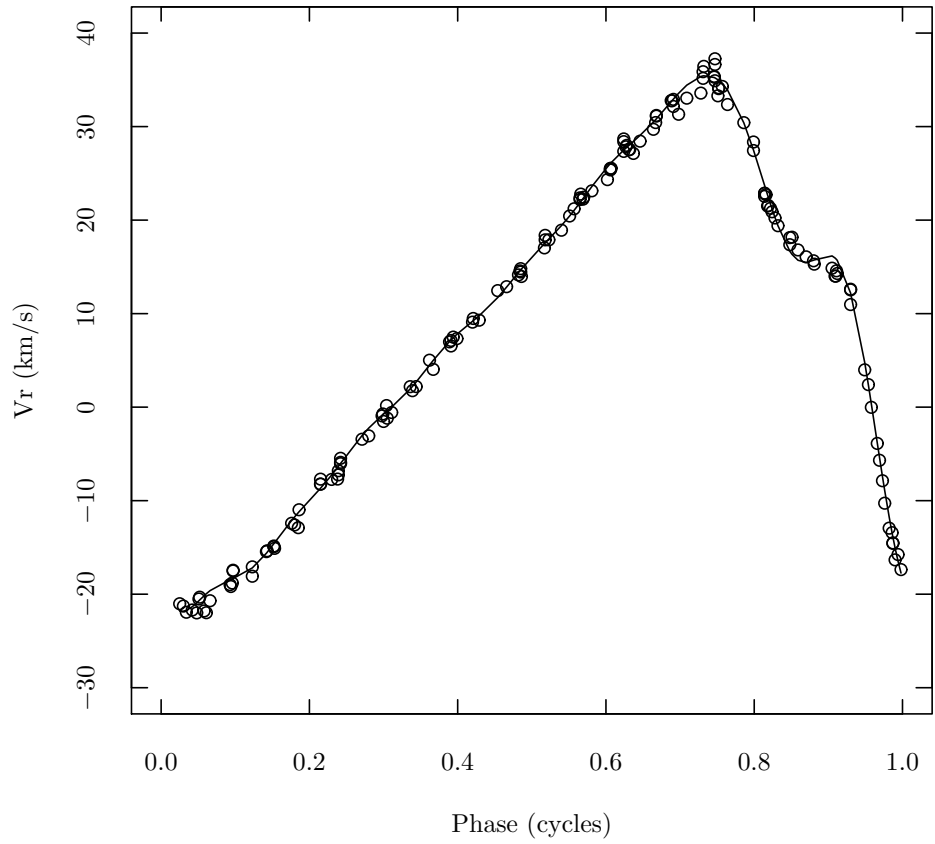}
\caption{The radial velocity curve for X Cyg with low weight data suppressed for clarity.  A Fourier
polynomial of order $M=9$ is shown fitted to the observations. \label{fig5}}
\end{figure}

\begin{figure}
\figurenum{6}
\plotone{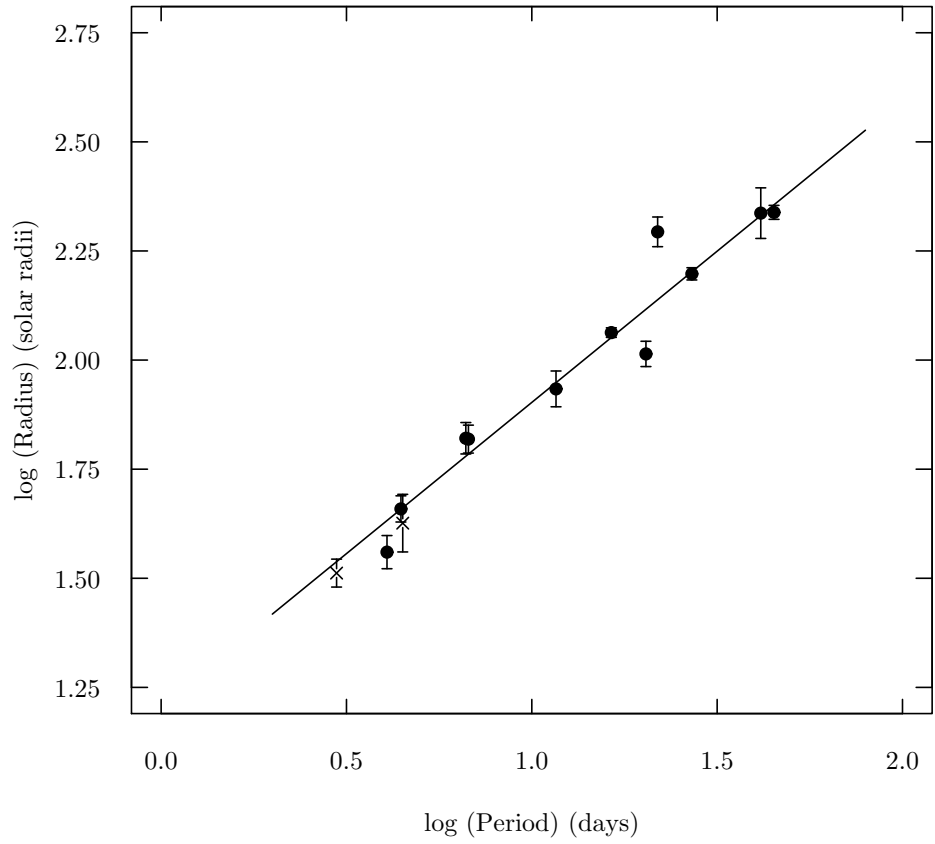}
\caption{Period--Radius relation for our sample. Fundamental-mode Cepheids are plotted with 
filled symbols; SZ Tau and EU Tau are plotted as X symbols at their computed fundamental periods. 
The line shows the weighted, least-squares fit. \label{fig6}}
\end{figure}

\begin{figure}
\figurenum{7}
\plotone{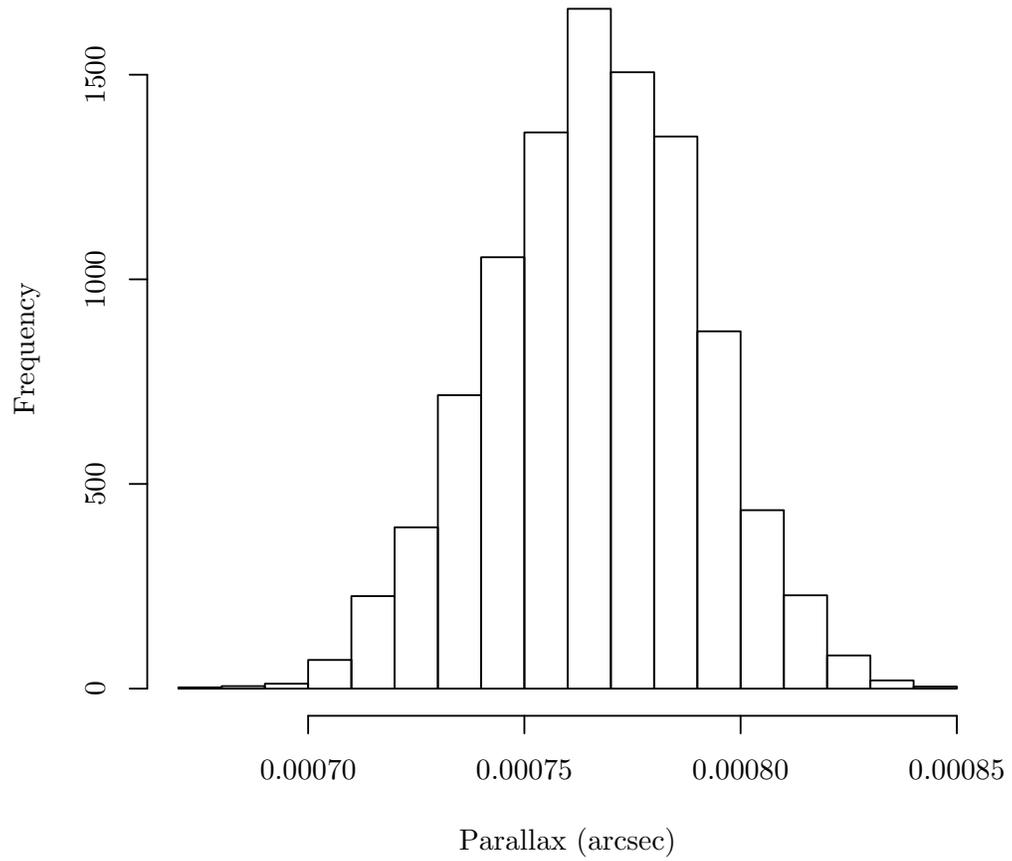}
\caption{Posterior marginal distribution of the parallax of T Mon. \label{fig7}}
\end{figure}

\begin{figure}
\figurenum{8}
\plotone{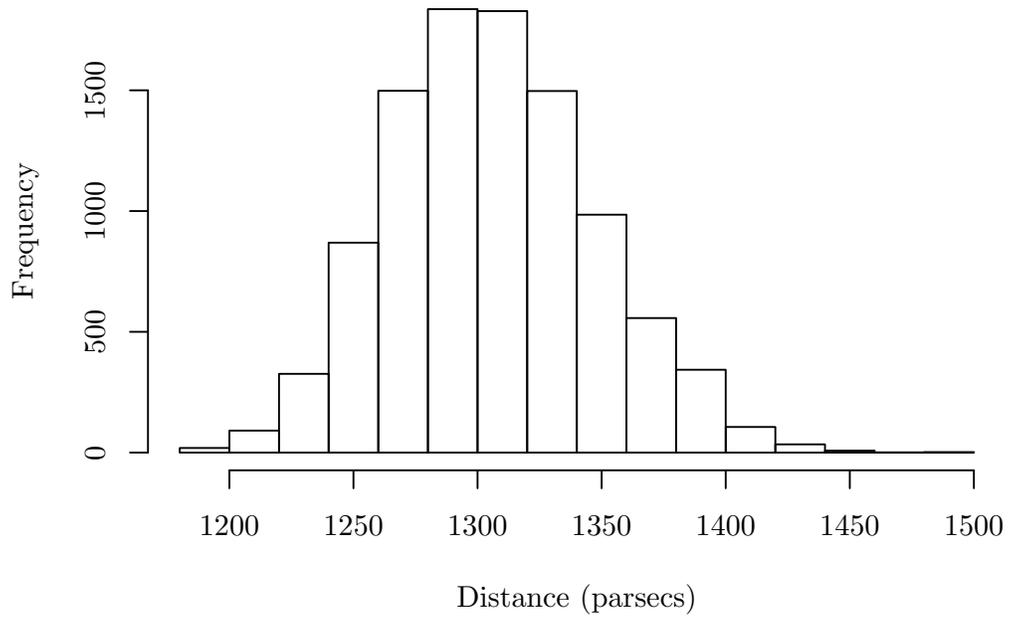}
\caption{Posterior marginal distribution of the distance of T Mon. \label{fig8}}
\end{figure}

\begin{figure}
\figurenum{9}
\plotone{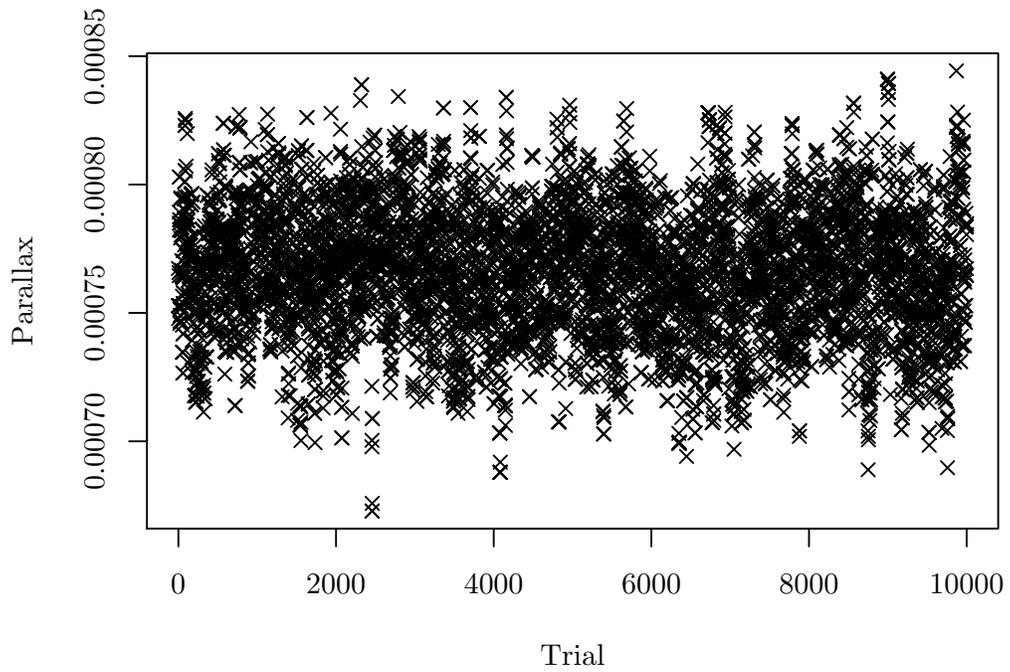}
\caption{Simulation history of the parallax of T Mon. \label{fig9}}
\end{figure}

\begin{figure}
\figurenum{10}
\plotone{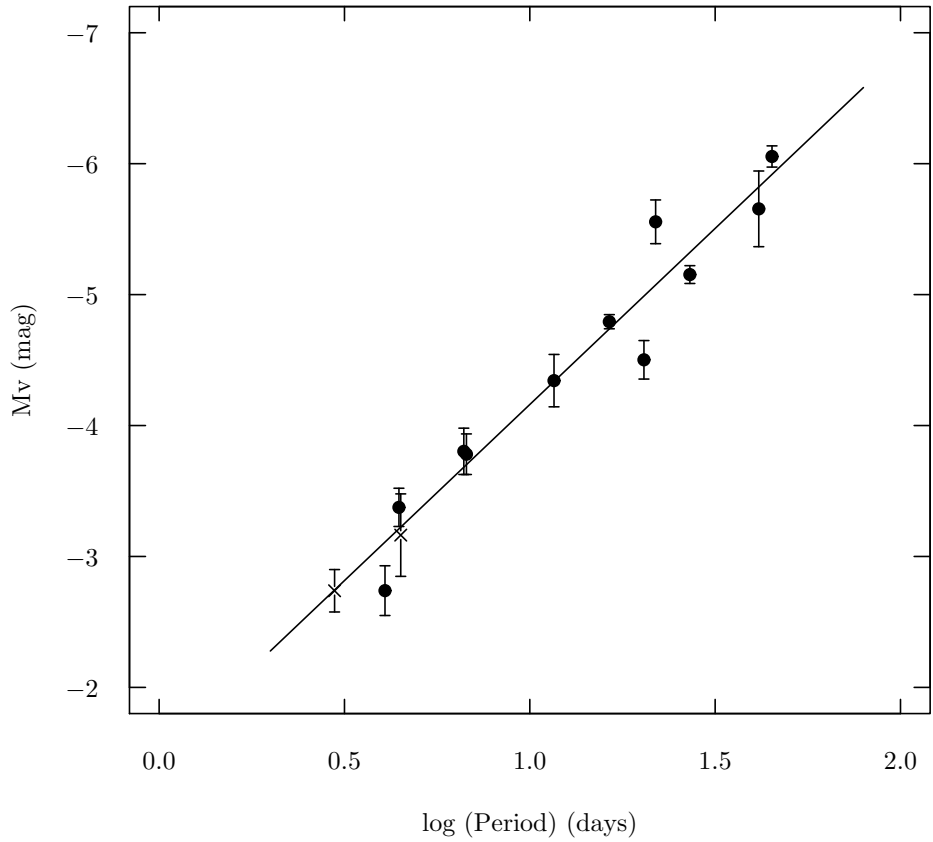}
\caption{Period-luminosity relation for our results.  The $<M_v(int)>$ magnitudes of Table 6
are plotted against log(period). Fundamental-mode Cepheids are plotted with filled symbols; SZ Tau and
EU Tau are plotted as X symbols at their computed fundamental periods.  The line shows the weighted,
least-squares fit.
\label{fig10}}
\end{figure}

\begin{figure}
\figurenum{11}
\plotone{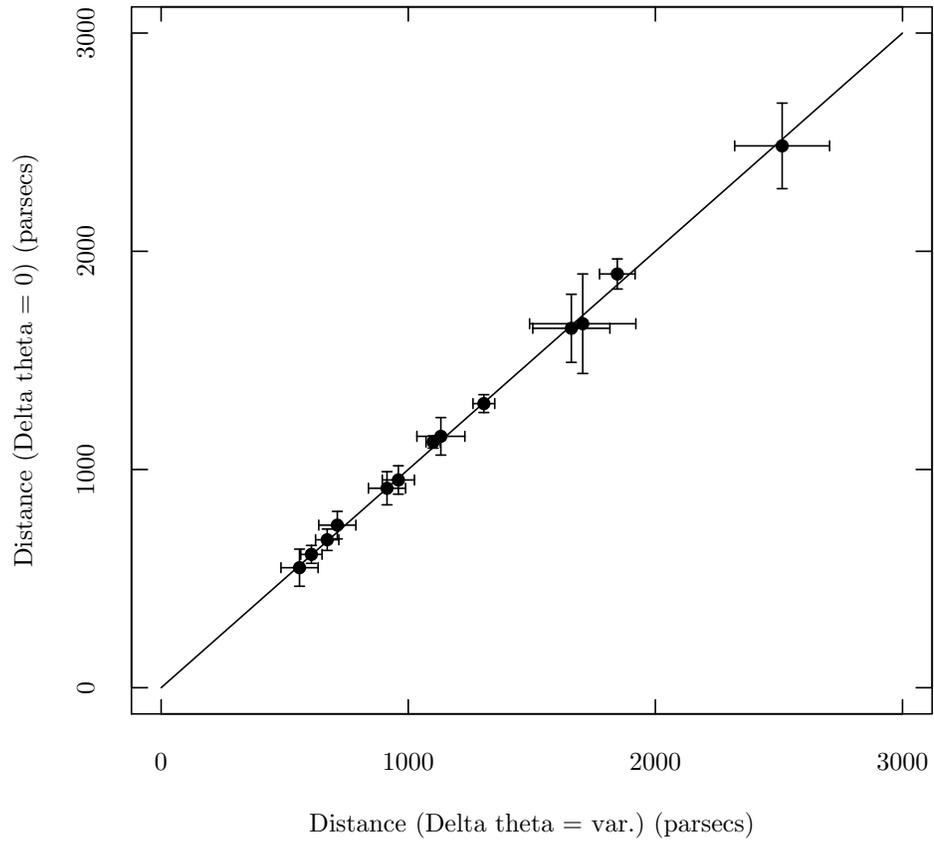}
\caption{Distances determined with $\Delta\theta$ fixed at zero plotted against distances obtained with 
$\Delta\theta$ variable. A line of equality is shown for comparison. \label{fig11}}
\end{figure}

\clearpage

\begin{deluxetable}{rrrl}
\tablenum{1}
\tablewidth{4.5in}
\tablecaption{Adopted Parameters}
\tablehead{\colhead{Cepheid} &\colhead{$E(B-V)$} &\colhead{Period} &\colhead{Period} \\&\colhead{(mag)}
&\colhead{(days)} &\colhead{Source}}
\startdata
RX Aur& 0.276& 11.623837& Moffett \& Barnes (1985)\\
X Cyg& 0.288& 16.385692& Szabados (1991)\\
T Mon& 0.209& 27.026& Evans {\em et al.} (1999)\\
BF Oph& 0.247& 4.067698& Szabados (1989)\\
RS Pup& 0.446& 41.467& This paper\\
U Sgr& 0.403& 6.745229& Szabados (1989)\\
WZ Sgr& 0.467& 21.849789& Szabados (1989)\\
BB Sgr& 0.284& 6.637005& Szabados (1989)\\
RY Sco& 0.777& 20.321044& Szabados (1989)\\
SZ Tau& 0.294& 3.14895& This paper\\
EU Tau& 0.172& 2.10248& Szabados (1977)\\
T Vul& 0.064& 4.435453& Szabados (1991)\\
SV Vul& 0.570& 45.019& This paper\\
\enddata
\end{deluxetable}

\begin{deluxetable}{rllll}
\tablenum{2}
\tablewidth{7in}
\tablecaption{Data Sources}
\tablehead{\colhead{Cepheid} &\colhead{Photometric} &\colhead{Photometric} &\colhead{Velocity}
&\colhead{Velocity} \\ &\colhead{Values} &\colhead{Sources} &\colhead{Values} &\colhead{Sources}}
\startdata  
RX Aur&  89&  7,12,13,14,19,35&                             58&  3,30,36               \\
X Cyg&  311&  1,4,7,8,9,10,12,13,14,18,35&                 199&  3,20,31,32,36         \\
T Mon&  181&  4,6,7,10,12,15,17,18,22,35&                  144&  3,20,23,32,36         \\
BF Oph& 183&  5,6,8,9,13,15,21,25,35&                       52&  3,24,33               \\
RS Pup&  81&  15,17,35&                                     26&  2                     \\
U Sgr&  330&  4,7,8,9,10,11,12,13,14,15,16,17,18,21,25,35& 156&  3,20,24,31,34,36      \\
WZ Sgr& 150&  7,12,13,14,22,35&                            104&  2,22,28,29            \\
BB Sgr& 281&  4,6,7,10,13,14,15,16,17,18,21,25,35&         117&  2,24,28,29,33         \\
RY Sco&  92&  15,22,35&                                     45&  2,22,33               \\
SZ Tau&  77&  1,35&                                        149&  3,20,27,28,29,31,     \\
\nodata& \nodata& \nodata&                             \nodata&  32,36                 \\
EU Tau& 345&  26&                                          118&  20,26,27,28,29        \\
T Vul&  305&  1,4,5,6,7,8,9,10,13,14,18,35&                190&  3,20,32,32,36         \\
SV Vul& 374&  1,4,5,6,7,8,10,11,12,13,14,18,35&            162&  3,20,30,32,36         \\
\enddata
\tablerefs{1--Barnes {\em et al.} 1997, 2--Barnes, Moffett, \& Slovak 1988, 3--Barnes, Moffett, \&
Slovak 1987, 4--Berdnikov 1993, 5--Berdnikov 1992a, 6--Berdnikov 1992b, 7--Berdnikov 1992c, 8--Berdnikov
1992d, 9--Berdnikov 1992e, 10--Berdnikov 1992f, 11--Berdnikov 1987, 12--Berdnikov 1986, 13--Berdnikov,
Ignatova, \& Vozyakova 1998, 14--Berdnikov, Ignatova, \& Vozyakova 1997, 15--Berdnikov \& Turner 2000,
16--Berdnikov \& Turner 1995a, 17--Berdnikov \& Turner 1995b, 18--Berdnikov \& Vozyakova 1995,
19--Berdnikov \& Yakubov 1993, 20--Bersier {\em et al.} 1994, 21--Caldwell {\em et al.} 2001, 22--Coulson
\& Caldwell 1985, 23--Evans {\em et al.} 1999, 24--Gieren 1981a, 25--Gieren 1981b, 26--Gieren {\em et
al.} 1989b, 27--Gorynya {\em et al.} 1992, 28--Gorynya {\em et al.} 1998, 29--Gorynya {\em et al.} 1996,
30--Imbert 1999, 31--Kiss 1998, 32--Kiss \& Vink{\'o} 2000, 33--Lloyd Evans 1980, 34--Mermilliod, Mayor,
\& Burki 1987, 35--Moffett \& Barnes 1984, 36--Wilson {\em et al.} 1989}
\end{deluxetable}

\begin{deluxetable}{rrrrrrrr}
\tablenum{3}
\tablewidth{7in}
\tablecaption{Diagnostic Results}
\tablehead{\colhead{Cepheid} &\colhead{$\Delta\theta$} &\colhead{$\sigma_{\Delta\theta}$} &N &M &$\sqrt{\tau_V}$ 
&$\sqrt{\tau_C}$ &$\sqrt{\tau_U}$\\ & & & Photometry & Velocities &$V$ &$(V-R)$ &Velocity}
\startdata
RX Aur&    $-0.0014$&  $\pm$0.0152&  4,5& 3& 0.82&  0.84&  0.82\\
 X Cyg&    $-0.0210$&  $\pm$0.0039&  9,10& 9,10,8& 0.68&  0.93&  0.57\\
	T Mon&    $ 0.0194$&  $\pm$0.0045&  6,7& 8& 0.59&  0.78&  0.55\\
BF Oph&    $-0.0489$&  $\pm$0.0128&  4& 3,2& 0.46&  0.58&  1.26\\
RS Pup&    $ 0.0102$&  $\pm$0.0199&  4& 3,4& 0.17&  0.56&  0.55\\
 U Sgr&    $-0.0294$&  $\pm$0.0111&  7& 5,7& 0.52&  0.73&  0.75\\
WZ Sgr&    $ 0.0237$&  $\pm$0.0127&  7,8& 7,8,9& 0.46&  0.56&  0.33\\
BB Sgr&    $-0.0190$&  $\pm$0.0122&  3& 4& 0.30&  0.63&  0.45\\
RY Sco&    $ 0.0135$&  $\pm$0.0100&  5,4& 1,2& 0.46&  0.89&  0.82\\
SZ Tau&    $-0.0031$&  $\pm$0.0219&  2& 2& 1.39&  0.68&  0.23\\
EU Tau&    $-0.0625$&  $\pm$0.0124&  3& 3,2& 0.53&  0.70&  0.51\\
 T Vul&    $-0.0215$&  $\pm$0.0103&  4& 4,5& 0.50&  0.72&  0.44\\
SV Vul&    $-0.0289$&  $\pm$0.0052&  6& 4,3& 0.33&  0.70&  0.13\\
Mean of 13&$-0.0115$&  $\pm$0.0065&   &  & 0.55$\pm$0.08&  0.72$\pm$0.03&  0.57$\pm$0.08\\
Mean of  7&$-0.0034$&  $\pm$0.0099&   &  &     &    &  \\
\enddata
\end{deluxetable}

\begin{deluxetable}{rrrrr}
\tablenum{4}
\tablewidth{4.5in}
\tablecaption{Radius Results}
\tablehead{\colhead{Cepheid} &\colhead{$<R>$} &\colhead{$\sigma_R$} 
&\colhead{$<\phi>$} &\colhead{$\sigma_\phi$} \\ &\colhead{Solar radii} 
&\colhead{Solar radii} &\colhead{milliarcsec} &\colhead{milliarcsec}}
\startdata
RX Aur&      85.9&  $\pm$8.1&  0.4514&  $\pm$0.0016\\
 X Cyg&     115.6&  $\pm$2.8&  0.9166&  $\pm$0.0017\\
	T Mon&     157.6&  $\pm$4.9&  1.0531&  $\pm$0.0022\\
BF Oph&      36.3&  $\pm$3.2&  0.4443&  $\pm$0.0013\\
RS Pup&     217.1& $\pm$29.0&  1.1106&  $\pm$0.0063\\
 U Sgr&      65.9&  $\pm$4.8&  0.8548&  $\pm$0.0020\\
WZ Sgr&     196.7& $\pm$15.3&  0.6831&  $\pm$0.0028\\
BB Sgr&      66.2&  $\pm$5.4&  0.6315&  $\pm$0.0014\\
RY Sco&     103.3&  $\pm$7.0&  0.9394&  $\pm$0.0032\\
SZ Tau&      42.3&  $\pm$6.4&  0.6593&  $\pm$0.0012\\
EU Tau&      32.5&  $\pm$2.4&  0.2507&  $\pm$0.0002\\
 T Vul&      45.6&  $\pm$3.1&  0.6547&  $\pm$0.0013\\
SV Vul&     218.0&  $\pm$8.1&  1.0306&  $\pm$0.0025\\
\enddata
\end{deluxetable}

\begin{deluxetable}{llrl}
\tablenum{5}
\label{allPR}
\tablewidth{6.5in}
\tablecaption{Cepheid Period--Radius Relations}
\tablehead{\colhead{Slope} &\colhead{Zero Point} &\colhead{Type} &\colhead{Source}}
\startdata
0.693$\pm0.037$ &2.042$\pm0.047$ &(V-R)        &This paper\\
0.737$\pm0.028$ &1.984$\pm0.025$ &KHG          &Turner \& Burke (2002)\\
0.649$\pm0.051$ &1.953$\pm0.042$ &(b-y)        &Arellano Ferro \& Rosenzweig (2000)\\
0.680$\pm0.017$ &1.962$\pm0.025$ &(V-R), (J-K) &Gieren, Moffett, \& Barnes (1999)\\
0.655$\pm0.006$ &1.974$\pm0.008$ &Theory$^*$   &Bono, Caputo, \& Marconi (1998)\\
0.750$\pm0.023$ &1.975$\pm0.028$ &(J-K)        &Gieren, Fouqu{\'{e}}, \& G{\'{o}}mez (1998)\\
0.610$\pm0.030$ &1.980$\pm0.03 $ &(B-V), (V-R) &Sachkov, Rastorguev, Samus, \& Gorynya (1998)\\
0.606$\pm0.037$ &1.990$\pm0.033$ &CORS         &Ripepi, Barone, Milano, \& Russo (1997)\\
0.741$\pm0.026$ &1.985$\pm0.031$ &(J-K)        &Laney \& Stobie (1995)\\
0.690$\pm0.018$ &1.979$\pm0.006$ & \nodata     &Weighted mean of nine solutions\\
& & & $^*$ models with $Z=0.02$, no convective overshoot\\
\enddata
\end{deluxetable}

\begin{deluxetable}{rrr}
\tablenum{6}
\tablewidth{4in}
\tablecaption{Overtone Cepheid Radii}
\tablehead{\colhead{} &\colhead{EU Tau} &\colhead{SZ Tau}}
\startdata
Period (days)&                 2.10248&          3.14895\\
$<R>$ Table 4&               32.5$\pm$2.4&     42.3$\pm 6.4$\\
$<R>^*$ no overshoot&          31.2&             42.2\\
$<R>^*$ with overshoot&        28.6&             38.6\\
$<R>$ from Eq. \ref{eq:PR}&    23.6&             31.2\\
$^*$ radii from Bono {\em et al.} (2001) \nocite{bono2001}
\enddata
\end{deluxetable}

\begin{deluxetable}{rrrrr}
\tablenum{7}
\tablewidth{5in}
\tablecaption{Distance Results}
\tablehead{\colhead{Cepheid} &\colhead{$s$} &\colhead{$\pi$} &\colhead{$<M_V(mag)>$} 
&\colhead{$<M_V(int)>$} \\ &\colhead{parsecs} &\colhead{milliarcsec} &\colhead{mag}
&\colhead{mag}}
\startdata
RX Aur&      1660&      0.6076&     $ -4.322$&       $-4.343$\\
      &  $\pm$156& $\pm$0.0553&  $\pm$0.200&   $\pm$0.200\\
 X Cyg&      1101&      0.9092&      $-4.744$&       $-4.793$\\
      &   $\pm$28&  $\pm$.0227&  $\pm$0.054&   $\pm$0.054\\
	T Mon&      1306&      0.7664&      $-5.100$&       $-5.153$\\
	     &   $\pm$41& $\pm$0.0241&  $\pm$0.068&   $\pm$0.068\\
BF Oph&       713&      1.4133&      $-2.718$&       $-2.739$\\
	     &   $\pm$63& $\pm$0.1226&  $\pm$0.190&   $\pm$0.190\\
RS Pup&      1706&      0.5966&      $-5.597$&       $-5.655$\\
	     &  $\pm$228& $\pm$0.0794&  $\pm$0.288&   $\pm$0.289\\
 U Sgr&       672&      1.4951&      $-3.756$&       $-3.781$\\
	     &   $\pm$49& $\pm$0.1050&  $\pm$0.155&   $\pm$0.155\\
WZ Sgr&      2513&      0.4003&      $-5.494$&       $-5.556$\\
	     &  $\pm$196& $\pm$0.0304&  $\pm$0.167&   $\pm$0.167\\
BB Sgr&       914&      1.1011&      $-3.786$&       $-3.803$\\
	     &   $\pm$76& $\pm$0.0883&  $\pm$0.176&   $\pm$0.176\\
RY Sco&       960&      1.0469&      $-4.466$&       $-4.502$\\
	     &   $\pm$65& $\pm$0.0712&  $\pm$0.147&   $\pm$0.147\\
SZ Tau&       560&      1.8239&      $-3.156$&       $-3.163$\\
	     &   $\pm$85& $\pm$0.2566&  $\pm$0.315&   $\pm$0.315\\
EU Tau&      1132&      0.8882&      $-2.732$&       $-2.738$\\
	     &   $\pm$86& $\pm$0.0659&  $\pm$0.162&   $\pm$0.162\\
 T Vul&       608&      1.6515&      $-3.354$&       $-3.375$\\
	     &   $\pm$41& $\pm$0.1109&  $\pm$0.146&   $\pm$0.146\\
SV Vul&      1846&      0.5425&      $-6.003$&       $-6.055$\\
	     &   $\pm$69& $\pm$0.0201&  $\pm$0.081&   $\pm$0.081\\
\enddata
\end{deluxetable}

\begin{deluxetable}{llrl}
\tablenum{8}
\label{allPR}
\tablewidth{6.5in}
\tablecaption{Cepheid Period--Luminosity Relations}
\tablehead{\colhead{Slope} &\colhead{Zero Point} &\colhead{Type} &\colhead{Source}}
\startdata
$-2.690\pm0.169$ &$-4.699\pm0.215$ &(V-R)          &This paper\\
$-2.694\pm0.138$ &$-4.657\pm0.132$ &KHG            &Turner \& Burke (2002)\\
$-2.810\pm0.060$ &$-4.820\pm0.090$ &Hipparcos $\pi$&Feast (1999)\\
$-3.037\pm0.138$ &$-4.665\pm0.164$ &(J-K)          &Gieren, Fouqu{\'{e}}, \& G{\'{o}}mez (1998)\\
$-2.986\pm0.094$ &$-4.954\pm0.095$ &(V-R)          &Gieren, Barnes, \& Moffett (1993)\\
$-2.851\pm0.056$ &$-4.812\pm0.058$ &\nodata        &Weighted mean of five solutions\\
\enddata
\end{deluxetable}

\begin{deluxetable}{llr}
\tablenum{9}
\label{delCep}
\tablewidth{5in}
\tablecaption{Absolute Magnitude of $\delta$ Cep}
\tablehead{\colhead{Source} &\colhead{$M_v$} &\colhead{Difference} \\ &\colhead{mag} &\colhead{mag}}
\startdata
Benedict {\em et al.} (2002)                &$-3.47\pm0.10$   &\nodata\\
Nordgren {\em at al.} (2002)                &$-3.55\pm0.08$   &$-0.08\pm0.13$\\
PL relation this paper, Eq. \ref{eq:PL2}    &$-3.43\pm0.10$   &$0.04\pm0.14$\\
Mean PL relation, Eq. \ref{eq:PL}           &$-3.47\pm0.06$   &$0.00\pm0.12$\\
Freedman {\em et al.} (2001)                &$-3.47\pm0.03$   &$0.00\pm0.10$\\
\enddata
\end{deluxetable}

\end{document}